# Opportunity-Normalized Residence-Workplace Matching and the Scale-Sensitive Structure of Urban Commuting

Mingzhi Xiao[1*], Yuki Takayama[1]

**Abstract:** Urban spatial structure is commonly evaluated through the spatial distribution of homes and jobs or through aggregate commuting outcomes. Yet these approaches do not reveal how the opportunities created by urban form are selectively transformed into actual residence-workplace connections. This study introduces opportunity-normalized residence-workplace matching by comparing observed commuting distance distributions with opportunity-based distributions constructed from all within-city residence-workplace pairs weighted by residential and employment mass. Using Output Area-level data for nine British cities, we show that realized pairings are systematically more concentrated at shorter distances than the urban opportunity structure alone would predict. After normalization, matching intensity declines with distance in a recurrent but heterogeneous pattern that is approximately linear in log-log space in many cities and can be summarized by a city-specific distance-decay coefficient. London further reveals that this regularity is scale-sensitive: a comparatively flattened citywide pattern separates into consistently negative but heterogeneous relationships across employment-centered subsystems. Supplementary evidence from New York and Chicago shows similar attenuation patterns. These findings identify realized residence-workplace matching as a distinct layer of urban structure and suggest that, in complex metropolitan systems, meaningful spatial regularities may reside in coherent matching fields rather than in aggregate city boundaries.

## 1. Introduction

Urban spatial structure is commonly described through the distribution of population, employment, land use, and infrastructure across space. Yet cities are not organized only by where these elements are located. They are also organized by how residential and employment locations are connected through everyday commuting. These realized residence–workplace linkages shape labor-market integration, center–periphery relations, and the practical functioning of urban space. In this sense, urban structure is not only a matter of spatial arrangement. It is also a matter of how homes and jobs are selectively connected through daily matching (Horner, 2004; Levinson, 1998).

This distinction matters because much of the existing literature evaluates urban structure through indicators that describe either spatial configuration or aggregate commuting outcomes, but not the

1 Institute of Science Tokyo, 2-12-1 W6-9, Ookayama, Meguro-ku, Tokyo 152-8552, Japan
* Corresponding author: xiao.m.1475@m.isct.ac.jp

realized relation between potential and actual residence–workplace connections itself. Research on jobs–housing balance has generated important insights into commuting burdens, spatial mismatch, and the uneven alignment of employment and residential distributions (Blumenberg & King, 2021, 2024; Blumenberg & Siddiq, 2023; Horner & Marion, 2009). Accessibility measures describe the reachability of opportunities across urban space (Levine, 1998; Levinson, 1998). Average commuting distance summarizes realized outcomes, while excess-commuting approaches compare observed commuting against optimized or counterfactual benchmarks (Hamilton & Röell, 1982; Kanaroglou et al., 2015; Small & Song, 1992; White, 1988). These approaches are highly valuable, but they leave under-specified a more basic question: how the distance structure of potential residence–workplace pairings differs from the distance structure of the pairings actually realized within cities.

A related body of work comes closer to this issue by emphasizing commuting possibilities, constrained choice sets, and the broader opportunity structure within which commuting is produced. Studies of commuting possibilities and commuting spectra have shown that observed commuting choices should be distinguished from the wider set of possible residence–workplace combinations distributed across distance bands (Charron, 2007; Yang & Ferreira Jr, 2008; Zhou et al., 2018). More broadly, urban and transport research has long cautioned against reading observed commuting outcomes as unconstrained preference, since realized behavior reflects accessibility conditions, housing-market constraints, search frictions, moving costs, institutional arrangements, and unequal opportunity sets (Kawabata & Shen, 2007; Levine, 1998; Ton et al., 2020; Van Ommeren et al., 1999). These insights make clear that commuting is not simply a behavioral expression of distance aversion. It is the realized outcome of selection within a structured urban opportunity field.

This logic has recently been sharpened by mobility research that separates the geometry of feasible origin–destination pairings from the realized movements that select among them. Boucherie et al. (2025) show that observed movement distances cannot be interpreted independently of the distance distribution implied by spatial opportunity itself. Their contribution clarifies a broader methodological point: before an observed distance distribution is interpreted as evidence of behavioral tendency or spatial organization, one must distinguish what the urban structure makes possible from what is actually realized. However, their empirical analyses focus on movements between same-type locations, including residential relocation and everyday mobility between places such as points of interest. Residence–workplace matching remains conceptually related but empirically unresolved, even though it represents one of the most fundamental ways in which cities connect heterogeneous locations through routine movement.

This paper addresses that gap by bringing the decoupling logic of opportunity structure and realized movement into the domain of intra-urban residence–workplace matching. Specifically, it compares the observed commuting distance distribution within each city with an opportunity-based distance distribution constructed from all within-city residence–workplace pairs weighted by residential and employment mass. The ratio between the two is interpreted as an opportunity-normalized realized matching intensity. This framing makes it possible to ask a sharper question than whether short-distance commuting is common. It asks whether, once the city's own opportunity structure is taken into account, realized residence–workplace matching exhibits a systematic distance-sensitive structure of its own.

The central argument of the paper is that it does. Observed commuting is not simply a passive reflection of the city's spatial opportunity geometry. Instead, cities appear to realize some distance-specific residence–workplace pairings more intensively than others, even after the joint distribution of homes and jobs is taken into account. This means that realized residence–workplace matching constitutes an additional layer of urban structure relative to the opportunity field generated by the city itself. In empirical terms, that organization can often be summarized by a clear distance-decay relation in opportunity-normalized matching intensity, and in many cities this relation is approximately linear in log–log space.

The contribution of the paper is threefold. First, it defines opportunity-normalized residence–workplace matching as a distinct analytical object. By separating the distance geometry of potential residence–workplace pairings from the distance structure of realized commuting links, the paper identifies a layer of realized urban organization that is not directly captured by jobs–housing balance, accessibility, average commuting distance, or excess-commuting benchmarks. Second, it shows that the attenuation of realized home–work matching is not merely a mechanical consequence of the city containing more short-distance opportunities. Even after the opportunity-based pair distribution is taken into account, realized matching intensity declines with distance in a recurrent but heterogeneous way. The resulting coefficient therefore summarizes the distance-sensitive realization of home–work connections relative to the city's own opportunity composition, rather than raw commuting distance alone. Third, the paper demonstrates that the meaningful scale of this regularity is not necessarily the city as a whole. The London decomposition shows that a flattened citywide curve may result from the superposition of multiple localized matching fields, and that clearer regularities can re-emerge once the system is decomposed into employment-centered subsystems. This shifts the interpretation of the coefficient from a simple city-level statistic toward a scale-contingent parameter of coherent residence–workplace matching fields.

Research on employment subcenters and polycentric urban form has long shown that major metropolitan systems cannot always be understood as single-center spatial fields (Lin et al., 2015; Modarres, 2011; Redfearn, 2007). More generally, regularities that appear weak or flattened at an aggregate scale may re-emerge once the system is decomposed into more internally coherent components (Mori et al., 2020). The implication for commuting is straightforward: a flatter citywide opportunity-normalized matching curve need not imply the absence of an underlying regularity. It may instead indicate that the chosen unit of analysis is too aggregate relative to the internal organization of the urban system. The relevant analytical object may therefore not always be the whole city, but the coherent matching field within which realized residence–workplace matching is actually organized.

Figure 1 summarizes this conceptual distinction. The spatial distribution of residences and jobs generates an urban opportunity field, represented as a distance-structured set of potential residence–workplace pairings. Observed commuting flows reflect only a selective subset of that field. By comparing realized matching with the opportunity-based pairing structure, the paper derives an opportunity-normalized measure of realized residence–workplace matching. This framework makes clear that the paper is not concerned with commuting distance alone. It is concerned with how actual home–work connections are organized relative to what the city's own spatial structure makes possible.

The empirical analysis uses Output Area-level data for British cities. The main comparative cases are London, Birmingham, and Manchester, while supplementary analyses extend the evidence to Leeds, Sheffield, Bradford, Liverpool, Bristol, and Cardiff. The study first compares the observed commuting distribution with the opportunity-based pair distribution implied by the joint distribution of residents and jobs. It then examines the opportunity-normalized matching pattern across distance and summarizes that pattern with an empirical distance-decay coefficient estimated at the city level. The main empirical finding is that realized residence–workplace pairings are systematically more concentrated at shorter distances than the urban opportunity structure alone would predict. After normalization, matching intensity declines with distance in a clear and comparable way, and in many cities this attenuation is close to linear in log–log space. A supplementary external validation using the New York and Chicago metropolitan areas is reported in the appendix to assess whether a similar opportunity-normalized attenuation pattern is also visible outside the British urban context.

At the same time, the paper shows that this regularity is heterogeneous rather than uniform, and that part of this heterogeneity reflects analytical scale rather than only cross-city difference. London provides the clearest example. In aggregate citywide form, London displays a declining but locally flatter opportunity-normalized pattern than Birmingham and Manchester. This paper argues that such aggregate flattening should not be interpreted as evidence against the broader regularity. Rather, it may reflect the superposition of multiple localized matching fields within a more spatially complex city. A center-based decomposition of London therefore serves not as a peripheral case study, but as a direct test of whether opportunity-normalized regularity becomes more visible once the urban system is represented at a more appropriate internal scale.

The broader significance of this argument lies in what the resulting coefficient can and cannot represent. It is not interpreted here as a normative target, a universal law, or a fully identified behavioral primitive. Instead, it is treated as an empirical summary of how a city organizes realized residence–workplace matching toward shorter distances relative to its own opportunity field. If such a structure is recurrent across cities, then it provides a useful comparative benchmark for describing realized urban integration and linking urban morphology to everyday home–work connections. Its contribution therefore lies less in claiming a timeless law than in making visible a previously under-specified layer of realized urban structure.

The remainder of the paper proceeds as follows. Section 2 reviews the literature on urban structure, jobs–housing balance, excess commuting, opportunity-based interpretations of spatial behavior, and the implications of polycentric urban structure for commuting regularities. Section 3 develops the conceptual framework of opportunity-normalized residence–workplace matching. Section 4 introduces the data and methods. Section 5 presents the empirical results, including the London decomposition. Section 6 discusses the broader implications of the identified coefficient, its interpretation boundaries, its possible relevance for urban structure modeling, and its planning significance. Section 7 concludes.

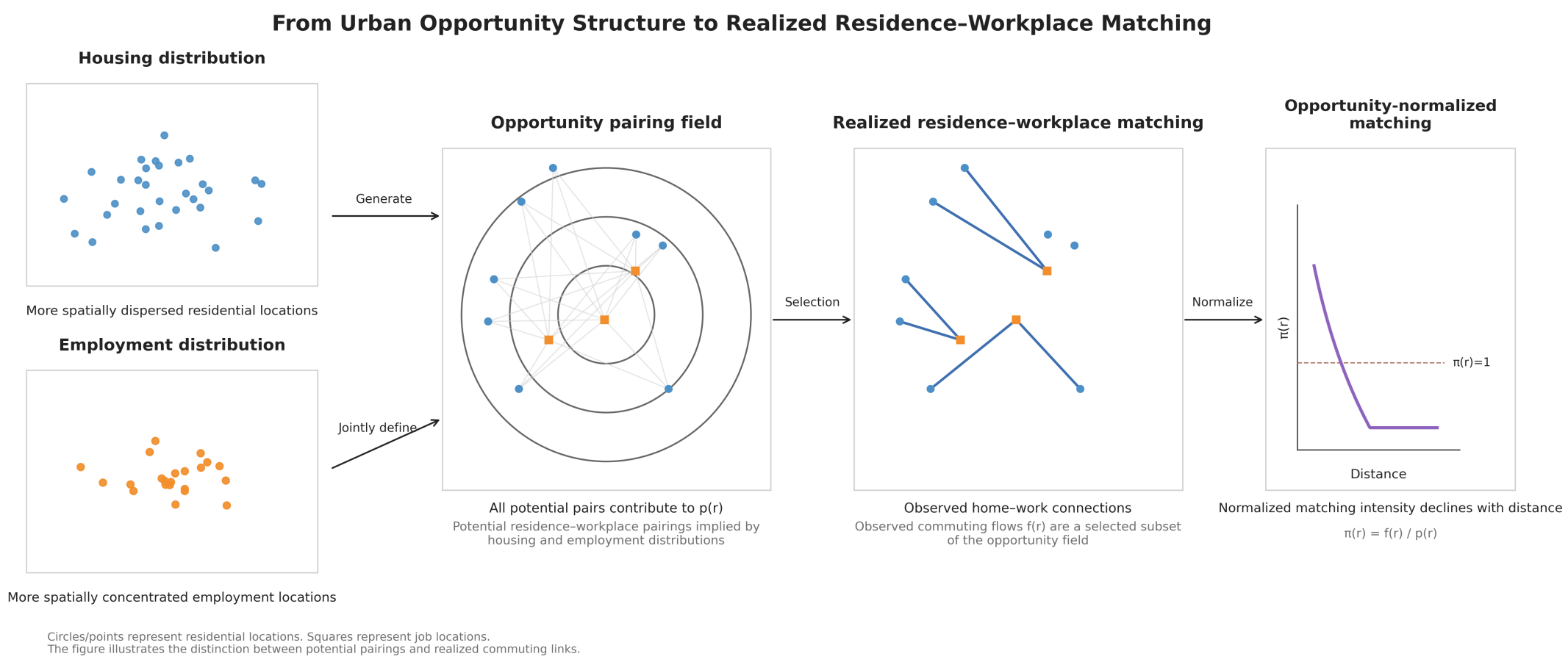

**Figure 1. Conceptual framework of opportunity-normalized residence–workplace matching**

# 2. Literature Review

## 2.1 Urban spatial structure as realized residence–employment organization

Urban spatial structure has long been studied through the spatial arrangement of housing, employment, centers, and flows. Classical and contemporary urban research alike emphasize that where people live and where economic activity is located are central to the internal organization of cities. Commuting occupies a special place within this broader field because it directly links residential and employment space. Levinson (1998) argued that journey-to-work patterns cannot be understood independently from accessibility conditions, while Horner (2004) emphasized that commuting should be interpreted as a multidimensional spatial phenomenon shaped by land use, labor markets, and transportation structure. Together, these studies suggest that commuting is not merely a transport outcome. It is also a spatial expression of how urban opportunities are arranged and connected.

Subsequent empirical work reinforces this point by showing that commuting outcomes reflect both urban structure and social heterogeneity. Antipova et al. (2011) found that urban land uses and socio-demographic attributes jointly influence commuting outcomes. Liu and Xiao (2023) showed that the relationship between built environment and commuting duration is nonlinear and differs between migrants and locals. O'Driscoll et al. (2024) demonstrated that socio-demographics, residential environments, and travel conditions interact in shaping commuter choices. Tong et al. (2024) argued that commuting responses depend on the built environment at both residence and workplace rather than on one-sided indicators alone. These studies collectively indicate that observed commuting patterns are shaped by the spatial organization of opportunity, by household and worker constraints, and by the way people select among available residential and employment combinations.

At the same time, much of this literature still treats commuting primarily as an outcome variable. The dominant focus remains on commute duration, commute distance, modal choice, accessibility, or transport conditions. These are all important, but they do not directly reveal how residential and employment distributions become linked through realized residence–workplace matching itself. If urban structure is partly produced through repeated home–work matching, then that matching process should be placed more centrally in urban analysis.

## 2.2 The limits of aggregate indicators

A major body of work has approached the relationship between housing, employment, and commuting through jobs–housing balance. This literature has shown that commuting burdens depend not only on the total number of jobs and residents but also on their relative distribution across space. Horner and Marion (2009), for example, proposed a spatial dissimilarity-based index and argued that jobs–housing balance should be interpreted spatially rather than through simple aggregate ratios. Subsequent studies showed that its relationship with commuting depends on urban form, distance decay, commuting mode, and spatial scale (Huang et al., 2021; Zheng et al., 2021; Zhou et al., 2018; Zhou et al., 2022).

Recent work has also emphasized that jobs–housing balance alone is insufficient. Blumenberg and King (2021) revisited the concept and argued that it remains useful but incomplete for understanding commuting outcomes. Blumenberg and King (2024) showed that jobs–housing balance is related to commute distance among young workers, but that this relationship remains mediated by housing-market conditions and regional context. Blumenberg and Siddiq (2023) further stressed jobs–housing fit rather than balance alone, highlighting that the relevant issue is not merely whether jobs and housing coexist in the same area, but whether the available housing is actually compatible with the workers employed there.

This literature establishes two important points. First, commuting should be evaluated against some spatial benchmark rather than treated as an isolated transport outcome. Second, aggregate indicators can conceal substantial internal heterogeneity. At the same time, jobs–housing balance remains primarily an indicator of configuration rather than realization. It tells us something about the relative arrangement of residents and jobs, but much less about which residence–workplace combinations are actually realized, and with what distance structure, once that arrangement is given.

The same limitation applies, in different ways, to accessibility indicators and average commute measures. Accessibility metrics capture potential reachability (Levine, 1998; Levinson, 1998). Average commute statistics summarize realized outcomes. But neither directly distinguishes between the urban field of potential pairings and the subset of pairings that are systematically realized. From the standpoint of urban structure, distinguishing potential pairings from realized matching is fundamental.

### 2.3 From minimum commuting to opportunity-based pairing structures

The excess commuting literature provides an important conceptual precursor because it explicitly compares observed commuting with a benchmark structure. Early studies argued that actual commuting exceeded the minimum feasible level, implying inefficiency or waste in urban organization (Hamilton & Röell, 1982). This interpretation was later challenged and refined. White (1988) showed that longer commuting patterns can emerge even under decentralized choice, while Small and Song (1992) demonstrated that the meaning of "excess" depends critically on how the benchmark is defined. Subsequent work further emphasized that both empirical estimates and policy conclusions are highly sensitive to the choice of counterfactual allocation and the spatial unit of analysis (Horner, 2002; Horner & Murray, 2002; Kanaroglou et al., 2015; Ling et al., 2024; Yang, 2008).

An especially important shift within this literature was the move away from optimized minimum benchmarks toward broader possibility-based interpretations of urban commuting structure. Charron (2007) argued that the more relevant question is not simply whether commuting exceeds some minimum, but how observed commuting relates to the wider field of commuting possibilities available within the city. Yang and Ferreira Jr (2008) developed this idea further through the commuting spectrum approach, explicitly distinguishing realized commuting choices from the broader set of potential job–housing combinations across distance bands. Zhou et al. (2018) extended this line of thinking with mobile phone data, examining jobs–housing balance and self-containment through commuting spectra rather than aggregate indicators alone.

This shift is fundamental for the present study because it redirects attention from optimal allocation to realized matching relative to the broader pairing field. Minimum-commuting approaches ask whether commuting is longer than necessary relative to an optimized benchmark. Possibility-based approaches ask how realized commuting is distributed relative to the set of pairings the city makes available. The present paper is aligned with this second logic. Its purpose is not to identify an efficiency gap or to recover a frictionless optimum. Rather, it asks how the realized distance structure of residence–workplace matching differs from the distance structure of all potential residence–workplace pairings generated by the city's observed residential and employment distributions.

A minimum benchmark is useful for evaluating whether observed commuting appears inefficient relative to a counterfactual allocation. An opportunity-based pairing benchmark, by contrast, makes it possible to evaluate how the city selectively realizes some distance-specific residence–workplace combinations more strongly than others. In this sense, the relevant contribution of the possibility literature is not merely that it offers another commuting benchmark. It establishes that the broader pairing field itself is analytically meaningful, and that commuting should be interpreted relative to that field rather than in isolation.

### 2.4 Opportunity structure, constrained realization, and the remaining gap

A broader literature in urban and transport studies reinforces the need for such a distinction by showing that realized commuting cannot be interpreted as unconstrained preference. Levine (1998) argued that accessibility and jobs–housing relations must be understood in relation to actual opportunity availability rather than simple locational proximity. Van Ommeren et al. (1999) showed that job location, residential location, and commuting emerge under search frictions and moving costs rather than under instantaneous optimization. Kawabata and Shen (2007) demonstrated that accessibility differs substantially across transport modes and population groups, while Ton et al. (2020) emphasized that observed commuting choices are shaped by experienced and constrained choice sets rather than by all theoretically available options.

Taken together, these studies make clear that observed commuting distances cannot be read directly as a pure behavioral signal. They are shaped by urban opportunity structure, institutional and market constraints, transport conditions, and selective realization processes. What remains less well specified, however, is how to represent this distinction empirically at the city level in a transparent, comparable way that is directly tied to residence–workplace matching itself.

Recent mobility research has sharpened this issue in a more general form. Boucherie et al. (2025) show that observed movement distances depend jointly on the geometry of feasible origin–destination pairings and the realized behavior that selects among them. Their contribution is important not only for mobility studies, but also for urban spatial analysis more broadly. It clarifies that an observed distance distribution cannot be interpreted on its own without first accounting for the distance distribution implied by opportunity structure itself. In other words, before one asks how people move, one must first ask what kinds of origin–destination pairings the spatial structure makes more or less available.

This point is directly relevant to commuting, yet an important gap remains. Existing studies rarely place the distinction between the full field of potential residence–workplace pairings and the subset of those pairings that is actually realized at the center of analysis. Jobs–housing balance focuses on configuration. Accessibility focuses on reachability. Excess commuting focuses on benchmarked outcomes. Commuting possibility studies come closer, but they still do not directly isolate realized residence–workplace matching relative to the city's own pairing field. As a result, a basic layer of realized urban organization remains under-specified.

The present paper addresses that gap directly. It compares the observed commuting distance distribution with an opportunity-based distance distribution constructed from all within-city residence–workplace pairs weighted by residential and employment mass, and interprets the ratio between them as an opportunity-normalized realized matching intensity. This is not intended to recover a pure behavioral preference parameter. Nor is it intended to represent the fully individualized feasible set of each commuter. Rather, it provides a transparent way to distinguish the geometry of potential pairings from the distance-specific intensity with which those pairings are actually realized. In this sense, the paper advances the literature not by replacing existing indicators, but by identifying a distinct analytical object that they do not directly capture.

### 2.5 Polycentric urban structure and the scale of regularity

A further strand of literature is essential for interpreting how such a relationship should be read in complex metropolitan systems. Large cities often contain multiple employment concentrations rather than one dominant center. Research on employment subcenters and polycentric urban structure has long shown that metropolitan spatial organization may consist of several partially distinct functional fields rather than a single homogeneous urban core (Lin et al., 2015; Modarres, 2011; Redfearn, 2007). This matters because citywide commuting patterns may reflect the superposition of several internally coherent residence–workplace systems rather than one unified matching field.

Empirical research on polycentricity and commuting has produced mixed conclusions about whether polycentric urban form necessarily reduces commuting burdens, but it consistently indicates that internal center structure matters for how commuting is organized (Lin et al., 2015; Modarres, 2011). For the present study, the key implication is not normative. It is analytical. If a large city contains multiple employment-centered subsystems, then a citywide opportunity-normalized matching curve may combine several distinct distance-decay relations into a single aggregate profile. In that case, apparent irregularity at the metropolitan scale may reflect aggregation across different matching fields rather than the absence of an underlying structure.

This concern connects to a broader theoretical insight about scale and apparent regularity. Mori et al. (2020) show that power-law structure may weaken or become blurred at more aggregate levels when multiple internally structured subsystems are combined, and may re-emerge after decomposition into more coherent units. A regularity that appears weak at the level of the whole system may still be present within its constituent subsystems if those subsystems operate at different effective scales or under different internal spatial logics.

This literature is central to the interpretation of London in the present study. If London's aggregate opportunity-normalized matching pattern appears flatter or less cleanly linear than those of Birmingham and Manchester, there are at least two possible readings. One is that the broader regularity identified elsewhere in the paper does not apply. The other is that the citywide unit is too aggregate, so that multiple localized matching fields are being combined into a single curve. The decomposition analysis is therefore not a peripheral extension or a descriptive side case. It is a direct test of whether the empirical visibility of opportunity-normalized regularity depends on internal urban scale.

This scale-sensitive interpretation is important because it broadens the meaning of heterogeneity in the paper. Differences in the estimated distance-decay relation may reflect not only differences between cities as whole units, but also differences in whether the chosen analytical scale corresponds to a coherent matching field. In simpler urban systems, the city may itself approximate the relevant functional unit. In more internally differentiated metropolitan systems, however, the meaningful unit of analysis may instead be the employment-centered subsystem within which realized residence–workplace matching is organized. The London case is used later in the paper precisely as an empirical test of this possibility.

# 3. Conceptual Framework: Residence–Workplace Matching and Urban Spatial Structure

Urban spatial structure is often described in terms of the spatial distribution of population, employment, land use, and infrastructure. Yet from the perspective of everyday urban functioning, these elements matter not only because of where they are located, but also because of how they are actually connected through repeated residence–workplace matching. A city is therefore not simply a static arrangement of residential areas and employment centers. It is also a system in which potential spatial relationships are selectively realized through daily commuting. From this perspective, commuting is not merely an outcome of urban structure. It is also a process through which urban structure becomes operational and visible.

Figure 1 summarizes the central conceptual distinction underlying the paper. The spatial arrangement of residences and jobs generates a broad field of potential residence–workplace pairings. Observed commuting flows, however, represent only a selectively realized subset of that broader opportunity structure. Building on this distinction, the present section clarifies the paper's analytical object and explains why the relation between potential pairing structure and realized matching structure matters for understanding urban spatial organization.

This distinction motivates the central conceptual move of the paper. Given the residential and employment distribution of a city, many residence–workplace combinations are in principle possible. However, only some of these combinations are actually realized in observed commuting flows. The realized commuting structure is therefore not identical to the full opportunity structure, because it reflects the selective realization of some residence–workplace combinations rather than others. The conceptual question addressed here is thus not simply how far people commute, but how actual residence–workplace matching is organized relative to the opportunity structure generated by the city itself.

This distinction is important because the observed commuting pattern alone can be misleading if interpreted without reference to the broader opportunity field. A high concentration of short-distance commuting may arise for at least two different reasons. First, the city may simply contain many more short-distance residence–workplace combinations in its underlying residential–employment geometry. Second, even given that opportunity structure, short-distance combinations may be disproportionately realized compared with longer-distance ones. The first concerns the spatial composition of potential pairings. The second concerns the relative intensity with which certain pairings are realized. Without separating them, it is difficult to know whether observed commuting distances mainly reflect opportunity structure or realized urban organization beyond it.

The conceptual framework of this paper therefore proceeds from a distinction between potential urban structure and realized urban structure. The potential structure refers to the full set of residence–workplace pairings implied by the spatial distribution of residents and jobs within the city. The realized structure refers to the subset of those pairings that appears in observed commuting flows. The difference between the two is analytically meaningful because it captures how the city's opportunity field is selectively translated into actual home–work matching. In this sense, the paper treats commuting as evidence of realized urban organization rather than as a stand-alone behavioral outcome.

This framing also helps position the paper relative to existing urban indicators. Measures such as jobs–housing balance, accessibility, and average commuting distance each capture important aspects of urban organization, but they do so from different angles. What these approaches do not directly isolate is the relation between the distance structure of potential pairings and the distance structure of realized matching. The present framework is designed to make that relation explicit.

A further reason this distinction matters is that realized commuting cannot be interpreted as unconstrained preference. A large literature has shown that commuting outcomes are shaped by search frictions, housing-market constraints, transport conditions, labor-market organization, and other institutional or spatial limitations (Kawabata & Shen, 2007; Levine, 1998; Ton et al., 2020; Van Ommeren et al., 1999). The framework proposed here is fully consistent with that view. It does not assume that observed commuting directly reveals pure behavioral preference. Instead, it asks a narrower and more tractable question: given the city's own opportunity structure, which distance-specific residence–workplace combinations are realized more or less strongly?

The opportunity baseline should therefore be interpreted carefully. In this paper, it is constructed from all within-city residence–workplace combinations weighted by residential and employment mass, and is deliberately parsimonious. It is not intended to represent a fully individualized feasible set, nor to incorporate every institutional, occupational, or demographic constraint that shapes actual matching. Rather, it provides a transparent city-level opportunity field generated by the existing joint spatial distribution of residences and jobs. Its analytical value lies precisely in this simplicity. Because the baseline does not absorb the full complexity of realized matching, the normalized pattern can be interpreted as the distance-sensitive structure of realization that is not mechanically implied by the city's opportunity geometry.

The opportunity-normalized matching measure developed later in the paper should therefore be interpreted as a system-level descriptive measure rather than as a direct indicator of individual preference. Even after normalization, the resulting pattern may still reflect transport accessibility, housing affordability, occupational sorting, search frictions, institutional arrangements, and other forces that are not explicitly modeled in the baseline. Its value lies in describing the distance-specific intensity with which residence–workplace combinations are realized relative to the city's own opportunity composition.

Seen in this way, the city can be understood as containing two analytically distinct but related layers: the spatial distribution of opportunities, and the realized organization of those opportunities through daily home–work matching. The core argument of the paper is that this second layer contains its own empirical structure. Once the opportunity field is taken into account, realized matching does not simply reproduce the geometry of available pairings. It may instead display a systematic distance-sensitive pattern, revealing how urban space is selectively organized through realized connection.

This point leads directly to the central conceptual object of the paper. If realized residence–workplace matching is evaluated relative to the city's own opportunity field, then the resulting distance-decay relation can be interpreted as a structural summary of how the city connects housing and employment space in practice. The paper's interest is therefore not limited to whether short-distance matching is common. It is whether the attenuation of opportunity-normalized realized matching can be summarized by a city-specific empirical coefficient. Such a coefficient would provide a compact description of how strongly a city organizes home–work matching toward shorter distances relative to its own opportunity structure, while allowing that this strength may vary across cities and spatial scales.

At the same time, the meaningful spatial unit of such a coefficient may itself depend on urban structure. In a relatively compact and weakly polycentric urban system, the citywide pattern may provide a reasonable approximation of a dominant matching field. In a larger and more internally differentiated city, however, the aggregate pattern may reflect the combination of several localized matching fields organized around different employment centers. In that case, a citywide coefficient may be harder to interpret because it summarizes a mixture rather than a single coherent subsystem. The London case examined later in the paper is important precisely because it allows this possibility to be tested empirically.

The implication is that a flattened aggregate pattern should not automatically be interpreted as evidence against a regular distance-decay structure. It may instead indicate that the relevant scale of regularity is smaller than the city as a whole. In this sense, the conceptual framework leaves open the possibility that the most meaningful unit for interpreting opportunity-normalized matching is not always the entire metropolitan area, but the internally coherent residence–employment matching field. This idea links the empirical analysis of London to a broader theoretical concern with scale-sensitive regularity in spatial systems.

To understand how commuting reflects urban spatial structure, one must compare the realized commuting distribution with the opportunity-based distribution implied by the city's own residential and employment structure. Only then is it possible to distinguish between the geometry of possible pairings and the selective realization of actual matching. The following section operationalizes this conceptual distinction by defining an observed commuting distribution, an opportunity-based pair distribution, and a ratio between them that captures opportunity-normalized realized matching intensity.

# 4. Data and Methods

### 4.1 Study areas, data, and analytical workflow

This study uses 2021 Output Area-level data for British cities. Output Areas are the smallest census-based statistical units consistently available across the study areas and are therefore suitable for fine-scale analysis of intra-urban residence–workplace structure. The commuting flows are derived from the 2021 Census origin–destination workplace flow data for England and Wales. The analysis begins with a large OA-level commuting dataset for England and Wales containing 10,741,185 observed OD records and a total commuting volume of 48,566,951. The empirical analysis reported here further restricts the sample to within-city residence–workplace flows in the nine study cities. This yields an analytical sample of 1,886,716 observed OD records and a total commuting volume of 2,553,224. For each OA, the analysis uses centroid coordinates together with residential and workplace counts. Observed commuting flows are represented as OA-to-OA links within the same city, while the opportunity structure is derived from the joint spatial distribution of residents and jobs across all OAs in that city.

The sample includes nine British cities: London, Birmingham, Manchester, Leeds, Sheffield, Bradford, Liverpool, Bristol, and Cardiff. London, Birmingham, and Manchester serve as the three main comparative cases because they provide clear contrasts in size, concentration, and internal spatial organization. The remaining six cities are used as supplementary cases to assess whether the main empirical pattern extends beyond the principal comparison.

For each city, the analysis retains only residence–workplace pairs located within the same urban boundary. City boundaries are defined consistently across the nine cases using urban boundaries constructed for this analysis, and OAs are assigned to each city according to whether their centroids fall within the corresponding boundary. The purpose is to focus on internal urban structure rather than inter-city commuting. Distances are measured as straight-line distances between OA centroids and grouped into 500-meter bins. In the main specification, same-OA pairs are excluded in order to reduce the mechanical concentration at zero distance and to focus on meaningful spatial separation within cities. This treatment avoids the mechanical dominance of zero-distance pairs and does not alter the substantive interpretation of normalized distance-decay beyond the immediate zero-distance bin. For external validation only, the appendix also reports metropolitan-scale results for the New York and Chicago MSAs using the same distributional logic of observed commuting, opportunity-based pair distributions, and opportunity-normalized matching intensity. These U.S. cases use 2022 block-level residence–workplace flow data from the LEHD Origin-Destination Employment Statistics (LODES) and Census-defined metropolitan boundaries. They are not used as part of the main British city comparison, but serve as supplementary evidence on whether the identified pattern is visible in a different national data environment.

Figure 2 summarizes the analytical workflow. The figure begins from two sources: observed OA-to-OA commuting flows and OA-level residential and employment counts. These are then used to construct the observed commuting distribution and the opportunity-based pair distribution, which are finally compared to derive an opportunity-normalized measure of realized matching and its distance-decay pattern across cities. In an additional London-specific decomposition reported in the Results section, the aggregate citywide pattern is further examined by identifying employment-centered subsystems and recalculating matching structures at the subsystem level. In this sense, the workflow does not merely produce descriptive distributions. It is designed to identify a city-level empirical summary of how realized residence–workplace matching attenuates with distance once the city's opportunity structure is taken into account.

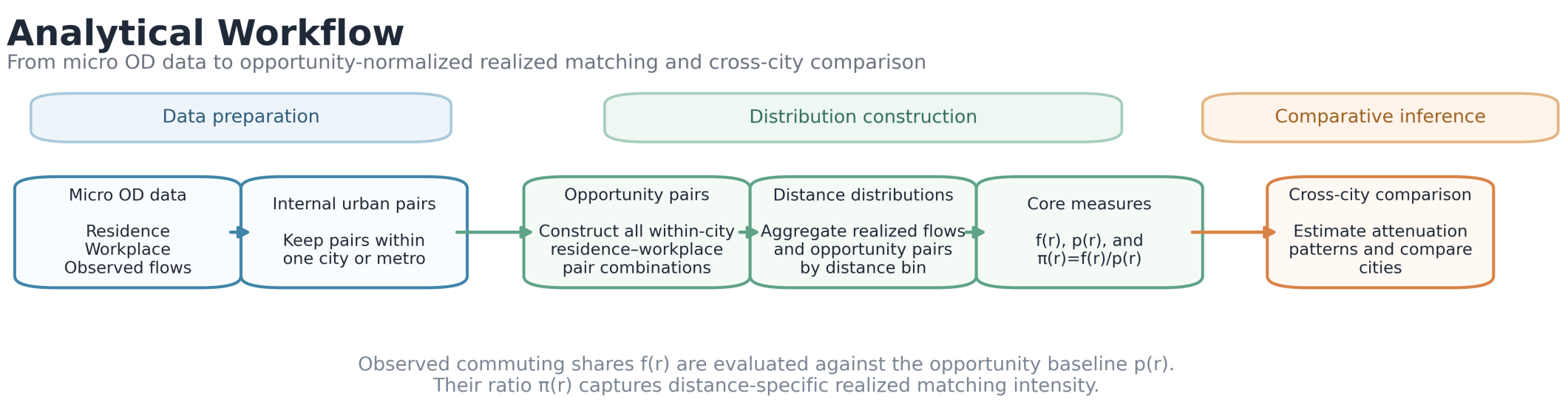

**Figure 2. Analytical workflow**

## 4.2 Observed and potential residence–workplace pairings

The analysis distinguishes between two distance distributions. The first is the observed commuting distribution, denoted by $f(r)$. This is constructed from actual within-city commuting flows between residential and workplace Output Areas (OAs). Each observed origin–destination pair is assigned to a distance bin according to the straight-line distance between OA centroids. Let $F_r$ denote the total observed commuting flow in distance bin $r$. The observed commuting share is defined as:

$$f(r) = \frac{F_r}{\sum_r F_r}$$

This distribution represents the share of realized within-city residence–workplace pairings that falls into each distance band.

The second is the opportunity-based pair distribution, denoted by $p(r)$. This distribution is constructed from all within-city residence–workplace OA pairs and captures the city's baseline pairing structure before realized selection is taken into account. Each pair is weighted by the product of the number of residents in the origin OA and the number of jobs in the destination OA. Let $P_r$ denote the total weighted number of within-city residence–workplace pairs in distance bin $r$. The corresponding share is defined as:

$$p(r) = \frac{P_r}{\sum_r P_r}$$

This distribution represents the distance structure implied by the city's joint residential and employment opportunity field.

The construction of $p(r)$ is deliberately parsimonious. Its purpose is not to reproduce the full individualized feasible set faced by each commuter, nor to incorporate every institutional, occupational, or demographic constraint that may shape actual residence–workplace matching. Instead, it provides a transparent city-level opportunity baseline generated by the observed joint spatial distribution of residents and jobs. In this sense, $p(r)$ should be understood as a structural opportunity field rather than a fully behavior-free abstraction.

A more complex baseline could, in principle, incorporate occupational matching, housing tenure, income segmentation, or transport network conditions. However, embedding such factors directly into the baseline would also absorb part of the very structure the present analysis seeks to examine. By keeping the baseline simple and comparable across cities, the framework allows the subsequent normalized relationship to be interpreted as a structural summary of how realized matching differs from the opportunity geometry implied by the city itself. The analytical objective is therefore not to construct the most behaviorally exhaustive benchmark, but to separate the spatial composition of potential pairings from their realized organization in observed commuting as clearly as possible.

It captures the distance composition of the full field of potential residence–workplace pairings under the city's observed residential and employment distribution. Its role is to provide a common baseline against which the realized commuting distribution can be compared in a transparent and structurally interpretable way.

### 4.3 Opportunity-normalized realized matching intensity

The central indicator of the analysis is the ratio between the realized commuting distribution and the opportunity-based pair distribution:

$$\pi(r) = \frac{f(r)}{p(r)}$$

This measure is interpreted as an opportunity-normalized realized matching intensity. Values above one indicate that residence–workplace combinations in a given distance band are realized more intensively than the city's opportunity structure alone would imply. Values below one indicate under-realization relative to that opportunity baseline. In this sense, $\pi(r)$ compares what the city makes possible with what it actually realizes.

The substantive importance of $\pi(r)$ is that it separates two analytically distinct reasons why short-distance commuting may be common. Nearby residence–workplace combinations may be common simply because the city contains many such combinations in its residential–employment geometry. But they may also be common because, given that opportunity structure, such combinations are realized more strongly than longer-distance ones. The ratio isolates this second dimension in a system-level descriptive sense. It therefore shifts attention away from raw commuting prevalence alone and toward the relative realization of distance-specific matching within the city's own opportunity field.

The opportunity baseline used here should nevertheless be interpreted carefully: $p(r)$ represents a city-level structural opportunity field generated by the existing joint spatial distribution of residents and jobs, not a frictionless individualized choice set. Importantly, $\pi(r)$ should not be interpreted as an individualized choice probability or a pure preference parameter. Even after normalization, the resulting pattern may still reflect transport accessibility, housing affordability, search frictions, tenure constraints, occupational matching, institutional arrangements, and other forces that are not explicitly modeled in $p(r)$ (Kawabata & Shen, 2007; Levine, 1998; Ton et al., 2020; Van Ommeren et al., 1999). What $\pi(r)$ captures is therefore more limited, but also more defensible: it describes the distance-specific intensity with which residence–workplace combinations are realized relative to the city's own opportunity composition.

This distinction is central to the paper's broader interpretation. The aim is not to explain all underlying mechanisms of commuting behavior, but to identify whether a systematic gap exists between the distance structure of potential pairings and the distance structure of realized matching. If such a gap is present, it indicates that the city selectively realizes some residence–workplace combinations more strongly than others. In that case, the distance profile of $\pi(r)$ can be interpreted as evidence of a broader layer of realized urban organization.

### 4.4 Distance-decay estimation

To summarize how opportunity-normalized realized matching changes with distance, the study examines the log–log relationship between $\pi(r)$ and $r$. Specifically, it estimates:

$$\log \pi(r) = \alpha + \beta \log r + \varepsilon_r$$

The coefficient $\beta$ is used as a compact empirical summary of how sharply opportunity-normalized matching intensity attenuates with distance within each case. In the language of the paper's broader argument, $\beta$ summarizes the dominant distance-sensitive structure of realized residence–workplace matching after opportunity normalization.

The use of a log–log specification is motivated by the empirical pattern observed across cities. If opportunity-normalized realized matching declines in an approximately linear way in log–log space, then the relationship can be summarized parsimoniously by a single coefficient without discarding its essential comparative meaning. This common descriptive form allows heterogeneity to be expressed through differences in the estimated slope, without assuming that all cities share the same underlying structure.

The interpretation of $\beta$ should nevertheless remain scale-sensitive. In relatively compact urban systems, the citywide pattern may approximate a single dominant residence–workplace matching field, in which case the city-level coefficient may serve as a meaningful summary of that overall field. In larger and more spatially differentiated systems, however, the aggregate citywide pattern may combine several internally distinct matching fields organized around different employment centers. In such cases, the estimated citywide $\beta$ may reflect a mixture of localized structures rather than a single coherent one. This is why the London decomposition later in the paper is analytically important. It tests whether an apparently flatter aggregate pattern reflects the absence of a regularity or the superposition of multiple internally coherent matching fields.

To reduce instability in the sparse tail, the fitting range is truncated after the first distance bin in which either $f(r)$ or $p(r)$ becomes zero. This rule is applied consistently across cities to preserve comparability. Conceptually, this truncation is intended to avoid allowing extremely sparse tails to dominate inference about the main matching structure. It also helps ensure that the estimated slope reflects the portion of the distribution in which realized matching and opportunity structure are both meaningfully present, rather than being disproportionately shaped by edge observations with little substantive weight.

The resulting coefficient should therefore be understood as a summary of the dominant realized matching structure over the empirically relevant support of the data. It is not intended to recover every local irregularity in the distance profile, nor to imply that the entire commuting distribution is perfectly characterized by one line. Its purpose is to provide a transparent and comparable empirical summary that captures the main attenuation pattern across cities and, later, across employment-centered subsystems within London.

As a robustness check, the paper also estimates the distance-decay relationship using a likelihood-based formulation reported in Section 5.7.4. In that specification, the observed commuting distribution is modeled as the opportunity-based pair distribution reweighted by a distance-decay term. This allows the main attenuation pattern to be evaluated without relying solely on the OLS fit of $log\pi(r)$ on $logr$.

### 4.5 Interpretation boundaries

Three interpretation boundaries should be emphasized. First, $\pi(r)$ is not a pure preference parameter. Even after normalization by the opportunity baseline, the resulting pattern may still reflect transport accessibility, housing affordability, search frictions, occupational sorting, institutional arrangements, and other forces that are not explicitly modeled in the baseline.

Second, the opportunity baseline is defined at the city level. It captures the distance structure of the urban opportunity field generated by the joint distribution of residents and jobs, but it does not recover the individualized feasible set faced by each commuter. The framework therefore identifies a structural relation estimated at the city level rather than a person-level behavioral mechanism.

Third, the analysis uses straight-line centroid distance rather than network distance or travel time. The results therefore describe realized matching over geometric separation rather than generalized commuting cost. The estimated relationship should accordingly be interpreted as a structural distance-decay pattern in Euclidean urban space.

Within these boundaries, the framework provides a transparent way to compare realized residence–workplace matching with the opportunity structure that makes such matching possible. It allows the resulting distance-decay coefficient to be interpreted as a structural benchmark of how homes and jobs are connected within the existing urban system. The coefficient is therefore best understood as an empirical summary of how realized residence–workplace matching is organized relative to the city's own opportunity composition, and as a useful reference for comparative urban analysis and the interpretation of urban spatial structure.

## 5. Results

### 5.1 The spatial opportunity field of residence and employment

The first step is to examine the spatial field within which residence–workplace pairings are generated. Figure 3 presents the residential and employment spatial intensity patterns for the three main study cities. Across all three cases, residential patterns are more spatially dispersed than employment patterns, while employment remains more concentrated around central areas. This contrast is most pronounced in London, where residential presence extends widely across the metropolitan area but employment remains more strongly clustered around the urban core. Birmingham and Manchester display the same general pattern, though on smaller spatial scales and with more compact internal structure (Levinson, 1998; Horner, 2004).

Figure 3 establishes the spatial background against which all later normalization is performed. If the residential field is broader and the employment field is more concentrated, then the city's opportunity structure is already uneven before any actual commuting occurs. In that sense, the geography of homes and jobs defines the baseline pairing structure against which realized matching must be evaluated. This is precisely why the later parameter should not be read as a free-standing commuting statistic, but as a summary of realized matching conditional on the spatial field from which pairings emerge.

Figure 4 summarizes these contrasts more compactly by comparing the distance between residential and employment centers, residential spatial dispersion, employment spatial dispersion, and the employment-to-resident dispersion ratio. London appears as the most spatially extensive case, while Birmingham and Manchester are more compact. The importance of Figure 4 is therefore not merely descriptive. It shows that the three cities differ in the internal geometry through which home and work opportunities are arranged, and that this geometry varies in ways that are likely to matter for the realized matching structure observed later (Horner & Marion, 2009; Huang et al., 2021; Zhou et al., 2022). This also helps explain why the normalized matching parameter should not be interpreted independently of broader urban morphology.

The supplementary cities show the same broad contrast between more dispersed residences and more concentrated employment, although with city-specific variation in strength and form. These additional patterns are reported in Appendix Figure A1 and indicate that the opportunity structure documented in the main cases is not confined to London, Birmingham, and Manchester.

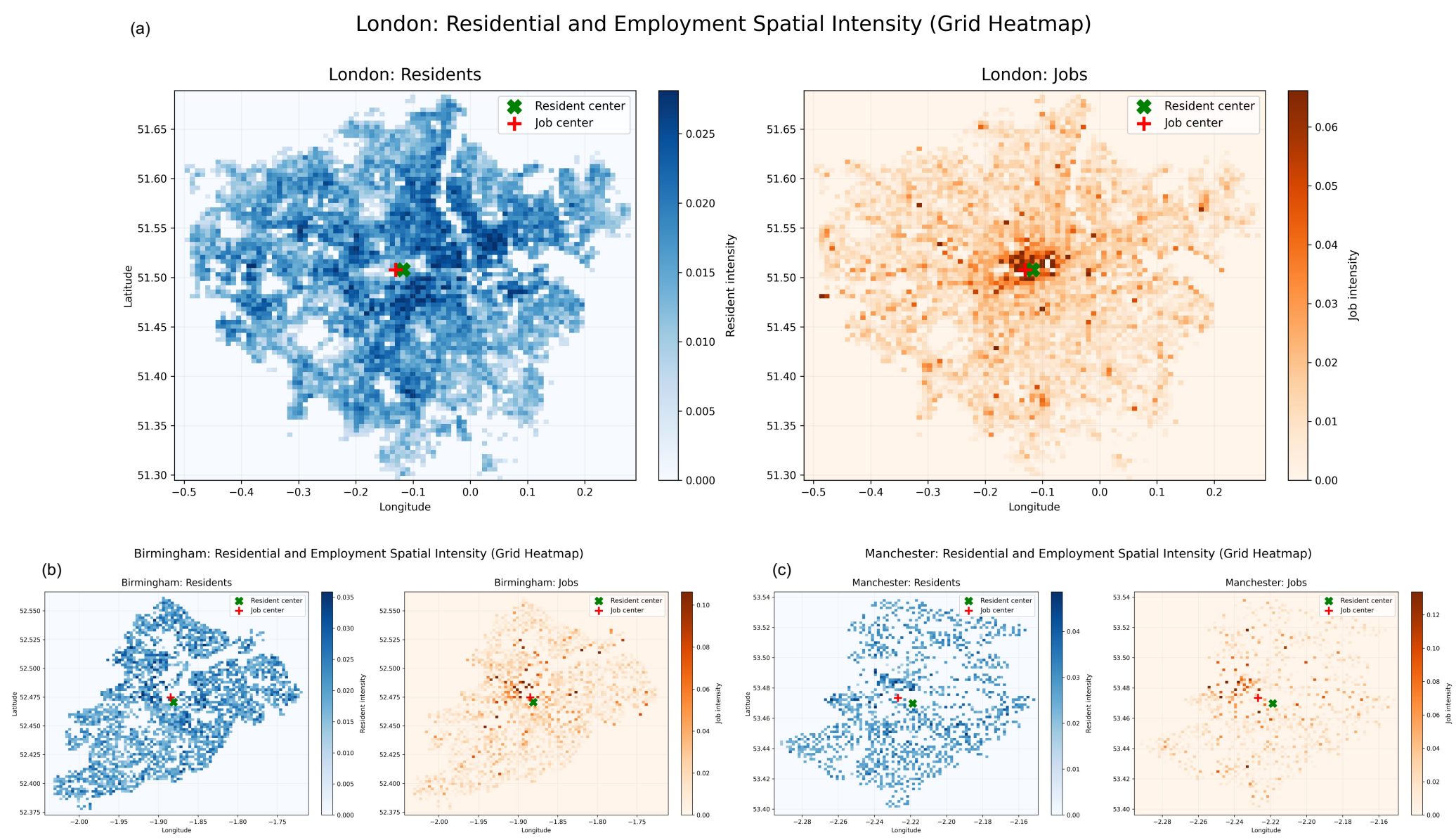

**Figure 3. Residential and employment spatial intensity in the main British study cities**

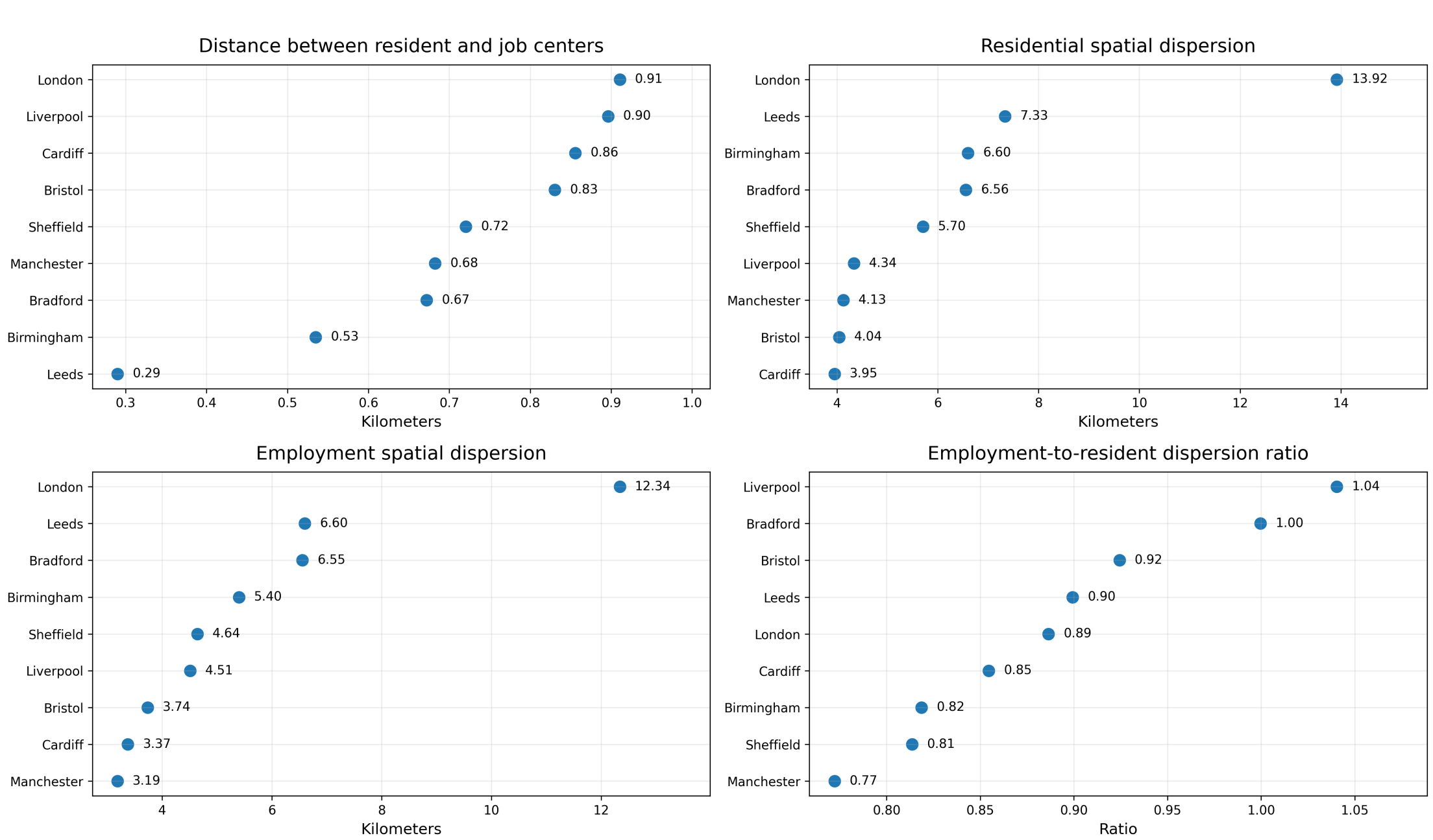

**Figure 4. Cross-city spatial summary metrics for residences and jobs**

## 5.2 Residential–employment structure and major commuting flows

Figure 5 turns from the opportunity field itself to the way dominant residential and employment areas are connected by actual commuting flows. Across all three main cities, employment-dominant grids are more concentrated than residential-dominant grids. The largest commuting flows connect these concentrated employment areas to a broader set of residential areas.

London exhibits the most extensive and multi-directional structure, with a dense employment core connected to a wide residential field. Birmingham and Manchester display the same basic logic, but at smaller scales and with more compact urban footprints. Figure 5 therefore provides a bridge between the morphological background in Figures 3 and 4 and the realized flow patterns that the later normalization seeks to interpret. In particular, the greater spatial complexity of London's major flows visually anticipates the more differentiated normalized pattern discussed later in the section on cross-city heterogeneity.

Comparable maps for the supplementary cities are reported in Appendix Figure A2. These appendix maps reinforce the idea that the relation between concentrated employment, more dispersed residence, and actual commuting flows is a recurring urban pattern rather than a city-specific anomaly. More broadly, Figure 5 makes clear that urban structure is not only about where jobs and residences are located in isolation, but also about the realized pathways through which they are linked. This point foreshadows the paper's central claim that the later parameter should be understood as a summary of structural integration rather than merely as a commuting slope.

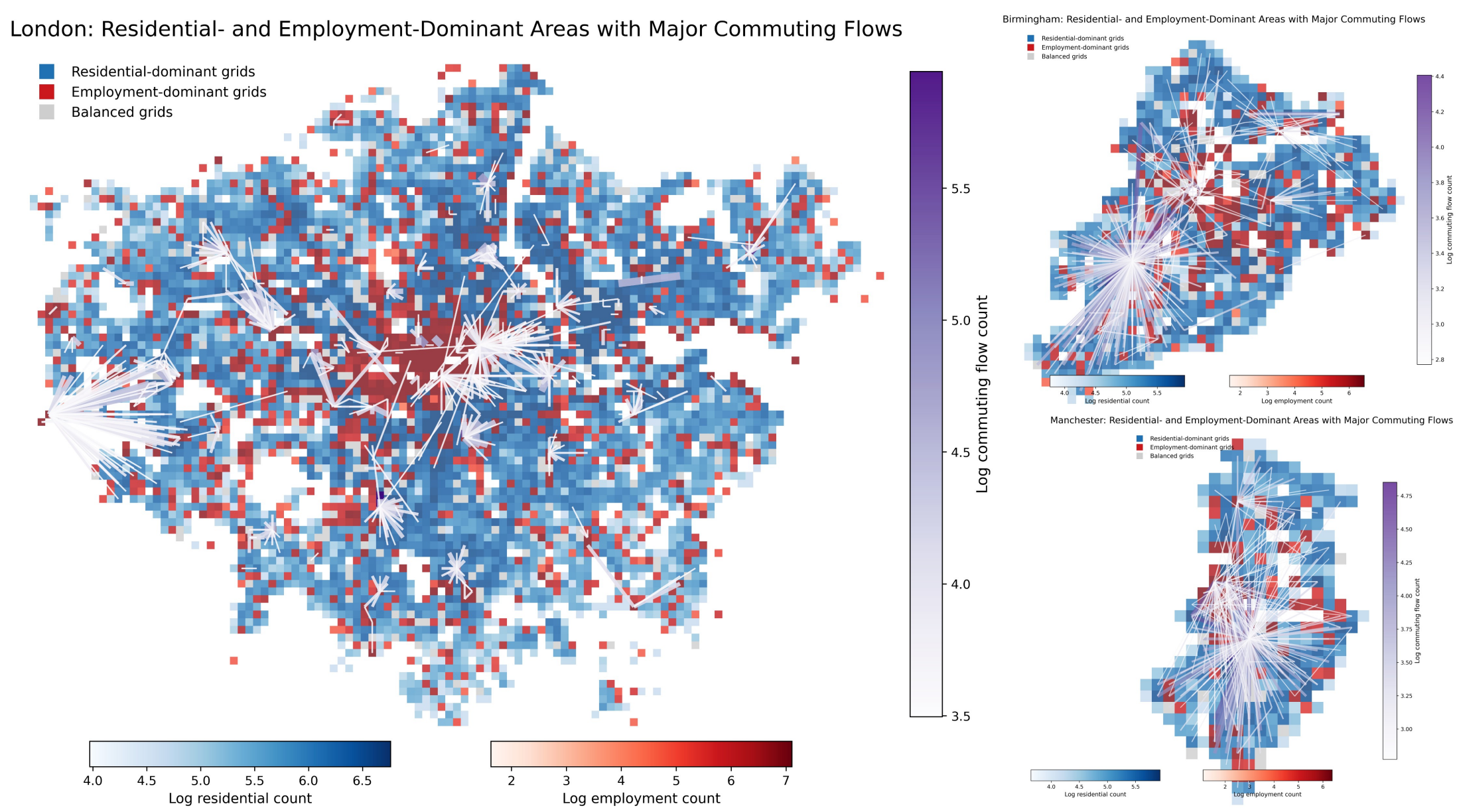

**Figure 5. Residential- and employment-dominant areas with major commuting flows in London, Birmingham, and Manchester**

## 5.3 Realized versus potential residence–workplace pairings

The core empirical comparison is presented in Figure 6, which places side by side the observed commuting distribution, $f(r)$, the opportunity-based pair distribution, $p(r)$, and their ratio, $\pi(r)$, for the three main study cities. This comparison is central because it makes it possible to distinguish two analytically different sources of short-distance concentration. One is purely geometric: a city may simply contain many more short-distance residence–workplace pairings in its underlying residential–employment structure. The other is selective realization: given that opportunity field, some distance-specific pairings may be realized more strongly than others.

Figure 6 shows clearly that the observed commuting distribution is more concentrated at shorter distances than the opportunity-based pair distribution in all three cities. By contrast, the opportunity-based distribution peaks at longer distances, reflecting the fact that the number of possible within-city residence–workplace pairings initially rises with spatial extent before eventually declining in the tail. This divergence is substantively important. It indicates that the commuting distance pattern actually realized within the city cannot be read directly from the city's opportunity geometry alone.

If short-distance commuting were common only because many short-distance pairings existed in the urban structure, the observed and opportunity-based distributions would be much more closely aligned. Instead, observed commuting is consistently shifted toward the shorter-distance range relative to the baseline pairing field. This means that cities do not simply contain potential residence–workplace pairings. They selectively realize some of those pairings more strongly than others. In that sense, the observed commuting distribution reflects not only the composition of opportunities, but also realized urban organization beyond it.

This comparison is fundamental to the interpretation of the later coefficient. The estimated slope does not summarize commuting distance in the raw sense. It summarizes the gap between the distance structure of all potential pairings and the distance structure of the pairings that are actually realized. The empirical distinctness of these two distributions is therefore what gives the later opportunity-normalized relationship substantive meaning. The right-hand diagnostic panels in Figure 6 are examined more directly in the next subsection.

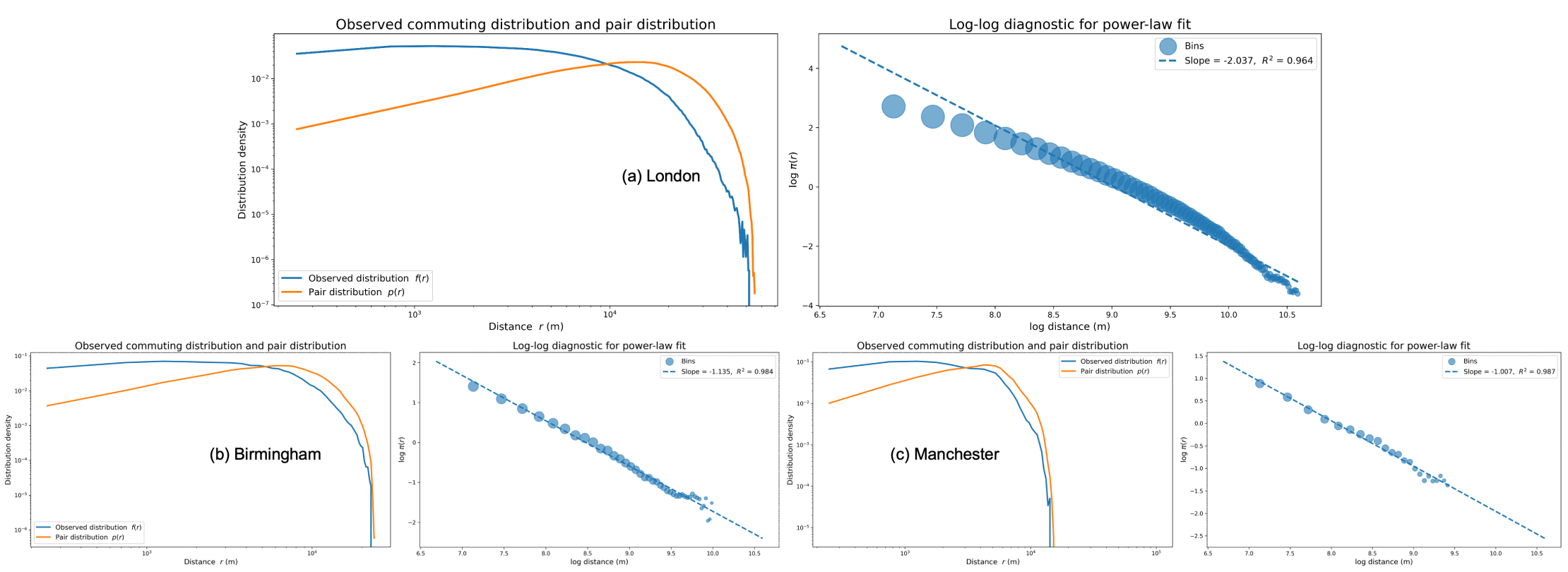

**Figure 6. Observed commuting distributions, opportunity-based pair distributions, and opportunity-normalized realized matching intensity in London, Birmingham, and Manchester**

### 5.4 Opportunity-normalized realized matching intensity and near-linearity

Once the observed commuting distribution is normalized by the opportunity-based pair distribution, a clearer structure becomes visible. Figure 6 shows that the resulting opportunity-normalized realized matching intensity declines systematically with distance in all three main cities. This means that short-distance residence–workplace combinations are not merely common in absolute terms. They are realized more intensively than longer-distance combinations even after the city's own opportunity structure is taken into account.

This is the central empirical regularity identified by the paper. It suggests that realized home–work matching is not simply a passive consequence of how residences and employment are arranged across urban space. Rather, it reflects an additional process of selective realization that consistently favors shorter-distance connections relative to the city's opportunity field.

The log–log diagnostics in Figure 6 suggest something stronger than monotonic decline alone. In each of the three main cases, the normalized relationship is close to linear over the main fitted range. This matters because it indicates that the distance-sensitive structure of realized residence–workplace matching can often be summarized parsimoniously by a compact empirical coefficient. The resulting pattern is therefore not just a generic short-distance bias. It exhibits a describable form that supports transparent comparison across cases.

The coefficient is therefore useful as a structural summary of how sharply realized home–work matching attenuates with distance relative to the city's opportunity composition. It does not describe commuting outcomes in isolation. Rather, it describes the dominant distance-sensitive form through which homes and jobs are functionally integrated once the opportunity field has been taken into account.

The supplementary cities reported in Appendix Figure A3 show the same broad qualitative pattern. In all cases, the observed commuting distribution is more short-distance concentrated than the opportunity baseline, and the opportunity-normalized matching intensity declines with distance. Taken together, these results suggest that the main regularity identified here is not confined to London, Birmingham, and Manchester alone. Rather, short-distance realized matching appears to be disproportionately strong relative to city-specific opportunity fields across a wider set of British urban contexts.

As an additional external validation, Appendix Figure A4 reports corresponding opportunity-normalized log–log diagnostics for the New York and Chicago metropolitan areas. Both U.S. cases exhibit strong negative and approximately linear attenuation patterns within 100 km. This additional evidence suggests that a similar regularity is not confined to the British urban sample, but can also be observed in large metropolitan systems with different spatial scales and urban forms.

### 5.5 Cross-city heterogeneity and why London requires decomposition

The presence of a common qualitative pattern does not imply that all cities share the same realized matching structure. On the contrary, one of the most important features of Figure 6 is that the steepness and local geometry of opportunity-normalized distance-decay differ across cities. In all three main cases, normalized matching declines with distance, but the attenuation is not equally sharp and the fitted relationships are not identical in local shape. The empirical regularity identified in the previous section is therefore recurrent but heterogeneous. Figure 6 supports two conclusions simultaneously: there is enough common structure for comparison, and enough variation to make that comparison meaningful.

This variation becomes more interpretable when Figure 6 is read together with the broader morphological background shown earlier in Figures 3–5. London is not only the largest city in the sample. It also displays the broadest residential dispersion, the broadest employment dispersion, and one of the largest separations between residential and employment centers. Birmingham and Manchester are more compact across all three dimensions. These contrasts matter because they suggest that the opportunity-normalized coefficient cannot be interpreted independently of urban internal structure. The same empirical form may summarize very different kinds of spatial organization depending on how residences, jobs, and centers are arranged within the city.

This issue is especially important for London. London's normalized relationship remains clearly declining, so the broader regularity identified in the paper is not absent there. At the same time, parts of the aggregate citywide curve are flatter and less cleanly linear than the simpler patterns observed in Birmingham and Manchester. Interpreted narrowly, this flatter curve might suggest that London follows a fundamentally different matching structure. But when read in light of its larger size, greater dispersion, and more complex employment geography, a more plausible interpretation emerges: the aggregate London pattern may reflect the superposition of multiple localized matching fields rather than a single coherent citywide one.

The analytical implication is important. London is not simply another case of cross-city heterogeneity. It raises the possibility that part of the apparent heterogeneity in opportunity-normalized matching arises from analytical scale. In a relatively compact city, the citywide unit may approximate a coherent matching field. In a more spatially differentiated metropolis, however, the citywide unit may mix together several internally structured residence–workplace systems. If so, the aggregate curve may appear flatter not because the underlying regularity disappears, but because several distinct matching fields are being combined into one profile.

This interpretation leads directly to a sharper empirical proposition. If the flatter aggregate London curve is a superposition effect, then decomposition into employment-centered subsystems should recover more interpretable localized opportunity-normalized distance-decay relations for centers with sufficient empirical support. The next section tests that proposition directly. In this sense, the London analysis is not introduced merely as an interesting case study. It functions as a test of whether opportunity-normalized regularity is itself scale-sensitive.

### 5.6 Decomposing London: internal scale and localized matching fields

To examine whether London's aggregate relationship reflects structural superposition rather than the absence of a broader regularity, this section decomposes the London commuting system into employment-centered subsystems. The purpose is not simply to provide a richer description of London. It is to test whether the opportunity-normalized distance-decay structure becomes more interpretable once the city is represented at a spatial scale that better corresponds to its internal employment geography.

Figure 7 shows the identification of employment centers in London. High-employment Output Areas are first identified from the upper tail of the employment distribution. Nearby high-employment OAs are then grouped into local peaks using a peak-separation rule, and commuting flows are assigned to employment-centered subsystems within the surrounding assignment radius. This procedure identifies 12 employment centers, each of which defines an employment-centered subsystem for the subsequent decomposition. These centers are not evenly distributed across the metropolitan area. Several are clustered around inner London, while others are located farther from the central employment concentration, including more peripheral employment nodes. This spatial pattern suggests that London is unlikely to be adequately represented as a single employment-centered field.

The decomposition is analytically important because a citywide opportunity-normalized curve may mix several partially distinct residence–workplace matching fields. If this is the case, the aggregate London pattern should not be interpreted as a simple exception to the broader distance-decay regularity. Instead, it may reflect the superposition of localized systems that operate at different spatial positions and effective scales. The center-level analysis therefore asks whether these localized subsystems reveal clearer and more interpretable opportunity-normalized matching relationships.

The center-level diagnostics are reported in Appendix Figure A6. Across the 12 employment-centered subsystems, the estimated slopes are consistently negative, ranging from −0.700 to −2.126. This indicates that, after normalization by the relevant opportunity structure, realized residence–workplace matching generally declines with distance within each employment-centered subsystem. The strength of the log–log fit varies across centers. The OLS $R^2$ values range from 0.803 to 0.953, with nine of the 12 centers exceeding 0.85 and seven exceeding 0.90. These results show that the decomposition does not produce one identical relationship across all centers. Rather, it reveals a set of localized matching fields that share a common negative distance-decay direction while differing in steepness and goodness of fit.

This heterogeneity is substantively meaningful. Centers located within or near the dominant inner-London employment structure tend to show relatively stable and interpretable attenuation patterns. By contrast, some peripheral or more spatially specialized centers display weaker linear fit or stronger curvature. This does not necessarily imply that these centers lack distance-sensitive matching. It more likely reflects the bounded, directional, and overlapping nature of commuting fields in a hierarchical polycentric metropolis. In such cases, the center-level matching field may be shaped not only by distance from the center, but also by the uneven geography of residential opportunities, transport corridors, and overlap with nearby employment nodes.

The decomposition therefore changes the interpretation of London. The aggregate London curve should not be read as evidence that opportunity-normalized distance decay is absent in large metropolitan systems. Instead, it is better understood as the combined outcome of multiple localized matching fields. Once these fields are examined separately, the negative distance-decay structure becomes visible across the full set of employment-centered subsystems, although its slope and local fit differ across subsystems. This supports the paper's broader argument that the empirical regularity of opportunity-normalized residence–workplace matching is scale-sensitive.

More broadly, the London case clarifies the meaning of heterogeneity in this paper. Cross-city differences do not arise only because cities differ as whole units. They may also arise because the analytical boundary does not always correspond to a coherent matching field. In a relatively compact city, the citywide unit may approximate the dominant residence–workplace matching field. In a complex polycentric city such as London, however, the citywide coefficient summarizes the combined effect of several partially overlapping subsystems. The most meaningful unit for interpreting opportunity-normalized matching may therefore be the coherent employment-centered matching field rather than the city as a whole.

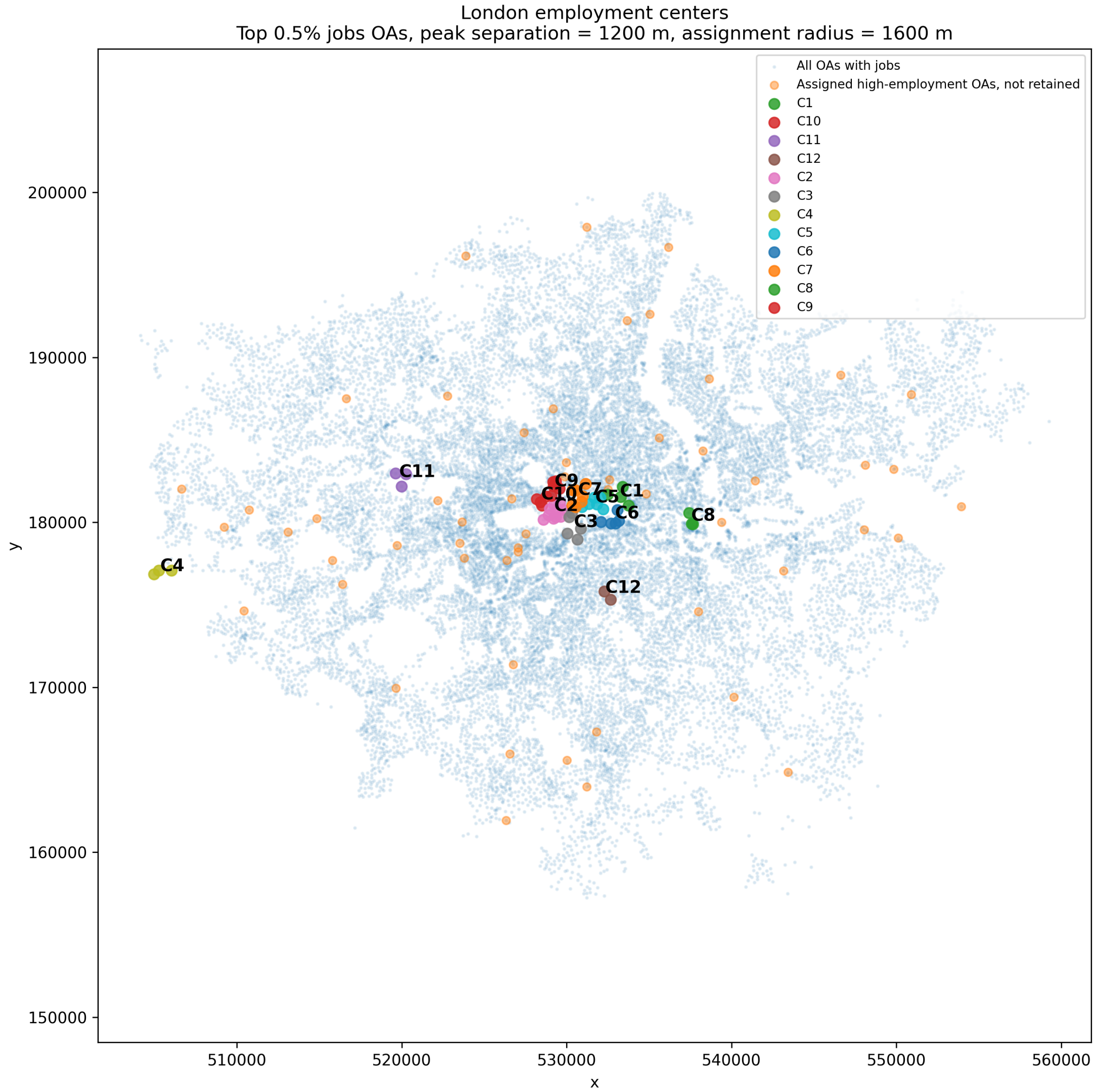

**Figure 7. Employment centers in London based on high-employment Output Areas, peak separation, and assignment radius**

**Notes:** Employment centers were identified using a rule-based procedure. High-employment Output Areas were first selected from the top 0.5% of the job distribution. Local employment peaks were then separated using a 1,200 m minimum peak-distance rule, and nearby high-employment OAs were assigned to the nearest peak within a 1,600 m radius. Centers were retained if they contained at least 8,000 jobs and at least two high-employment OAs.

## 5.7 Robustness

### 5.7.1 Robustness of city-level estimates under alternative truncation rules

While Section 5.6 refines the interpretation of the London case by decomposing the aggregate pattern, it is also important to examine whether the core city-level estimates remain stable under alternative truncation rules. Figure 8 compares the baseline specification with several alternatives, including no trimming, truncation at the first zero in the observed distribution, truncation at the first zero in the opportunity-based distribution, truncation at the first zero in either distribution, and fixed maximum-distance thresholds.

Figure 8 shows that the broad conclusions remain stable. London consistently exhibits a clear negative opportunity-normalized distance-decay pattern in the overall fitted relationship, while Birmingham and Manchester retain the same broad ordering and remain qualitatively similar across specifications. Although the exact slope values vary somewhat, all cities continue to display a negative attenuation pattern in opportunity-normalized matching intensity.

This stability matters because it indicates that the main findings are not sensitive to how the sparse tail is treated. The short-distance concentration of realized matching relative to the opportunity baseline, and the broader near-linear decay structure of the normalized relationship, remain visible across alternative truncation rules. The coefficient is therefore not perfectly invariant, but its substantive interpretation as a structural summary of distance-sensitive home–work integration remains intact across reasonable alternative specifications.

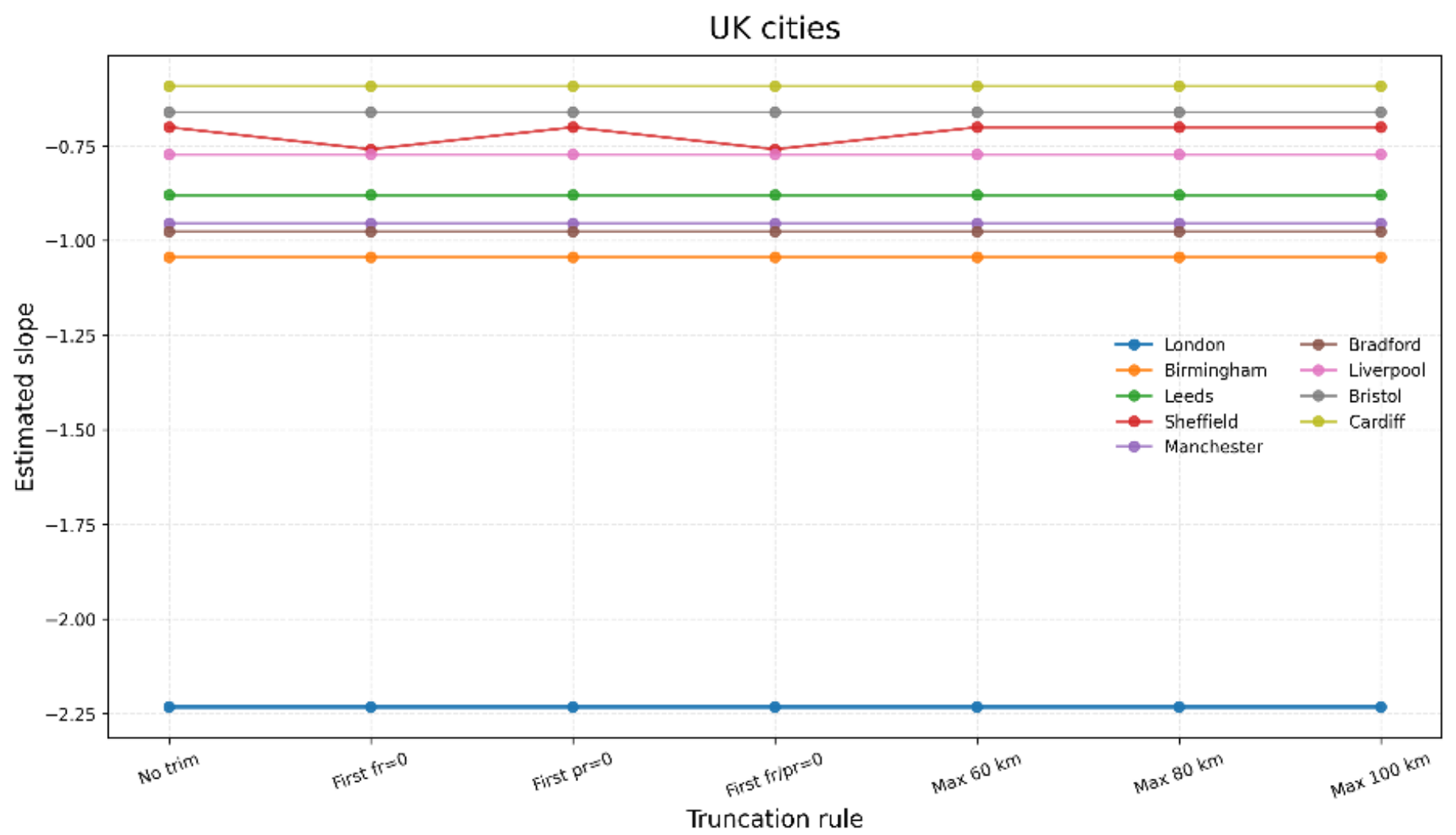

**Figure 8. Robustness of estimated distance-decay slopes under alternative truncation rules**

### 5.7.2 Robustness to alternative distance bin widths

Because the main analysis uses 500-meter distance bins, it is important to examine whether the estimated opportunity-normalized distance-decay pattern is sensitive to this bin-width choice. As an additional robustness check, the London city-level analysis was re-estimated using 250-meter and 1-kilometer bins after excluding zero-distance observed flows and same-OA opportunity pairs in a manner consistent with the main specification. Appendix Figure A5 shows that the main qualitative result remains unchanged. In both alternative specifications, opportunity-normalized realized matching continues to decline approximately linearly in log–log space, and the estimated slopes remain close to the baseline magnitude. These results indicate that the London city-level pattern is not an artifact of the specific 500-meter bin width used in the main analysis.

### 5.7.3 Robustness of the London decomposition

To assess whether the London decomposition depends heavily on weak empirical support in sparse distance bins, an additional robustness check was conducted by varying the minimum observed-flow threshold required for distance bins included in the fitted support. This check focuses on whether the center-level distance-decay results remain stable when very low-flow bins are excluded from the estimation range.

The results are reported in Appendix Table A3. Under minimum bin-flow thresholds of 5, 10, and 20 commuters per bin, all 12 employment-centered subsystems are retained. The mean OLS slope remains highly stable across the three thresholds, changing only from −1.106 to −1.080. The mean OLS $R^2$ also remains high, ranging from 0.931 to 0.943. Similarly, the mean MLE-implied slope remains close across specifications, ranging from −1.023 to −1.015, while the mean correlation between the observed commuting share and the model-implied share remains above 0.95 in all cases.

These results indicate that the center-level findings are not driven by isolated low-flow distance bins or by instability in the sparse tail of the distribution. The full set of 12 employment-centered subsystems remains empirically supported under increasingly restrictive bin-flow thresholds, and the estimated attenuation patterns remain substantively similar. The London decomposition should therefore be interpreted as a stable representation of localized residence–workplace matching fields, rather than as a fragile result of one particular fitted support.

Appendix Table A2 reports the identification settings used to define the London employment centers, including the high-employment OA definition, peak-separation distance, assignment radius, distance bin width, and minimum starting distance. Appendix Table A3 then shows that the resulting center-level estimates remain stable under alternative minimum bin-flow thresholds.

**5.7.4 Likelihood-based validation of bounded distance-decay patterns**

As a further robustness check, the distance-decay relationship was re-estimated using a likelihood-based formulation rather than relying only on the OLS log–log diagnostic. The purpose of this check is to assess whether the main attenuation pattern remains visible when the observed commuting distribution is modeled directly as an opportunity-conditioned distribution. Because the main analysis is defined over binned distance distributions, this validation is also implemented at the distance-bin level rather than at the individual OD-record level. The resulting estimates should therefore be interpreted as distribution-level validation of bounded distance decay, rather than as individual-level continuous power-law estimation.

The main likelihood-based specification defines the model-implied commuting share in distance bin $r$ as:

$$q(r;s) = \frac{p(r)r^{-s}}{\sum_k p(k)k^{-s}},$$

where $p(r)$ is the opportunity-based pair distribution and $s$ is the distance-decay exponent. The exponent is estimated by maximizing the multinomial log-likelihood:

$$\ell(s) = \sum_r F_r \log q(r;s),$$

where $F_r$ is the observed commuting flow in distance bin $r$. For comparability with the OLS log–log diagnostic, the MLE-implied slope is reported as $-s$.

Because intra-urban commuting distances are geographically bounded, the pure opportunity-conditioned power-law model was also compared with three alternative specifications: an opportunity-only baseline, $q(r) = p(r)$; an opportunity-conditioned exponential decay model, $q(r;\lambda) \propto p(r)e^{-\lambda r}$; and a bounded or truncated power-law model, $q(r;s,\lambda) \propto p(r)r^{-s}e^{-\lambda r}$. Model fit was evaluated using the Kolmogorov–Smirnov statistic, the correlation between the observed commuting share $f(r)$ and the model-implied share $q(r)$, and information criteria.

Table 1 reports the main likelihood-based validation results for the 12 employment-centered subsystems in London. Appendix Table A5 reports the empirical support of the binned likelihood-based estimates, including the fitted distance range, number of positive distance bins, observed commuting flow, exponent, and MLE-implied slope for each employment-centered subsystem. The results are consistent with the OLS diagnostics reported in Appendix Figure A6. All OLS slopes are negative, ranging from −0.700 to −2.126, and all MLE-implied slopes are also negative, ranging from −0.728 to

−1.833. This directional consistency is important. It shows that the negative opportunity-normalized distance-decay pattern is not an artifact of fitting a straight line to $log\pi(r)$. The same basic attenuation structure remains visible when observed commuting shares are modeled directly as an opportunity-conditioned distribution.

The MLE-implied slopes are generally less steep than, or close to, the OLS slopes. This difference is expected because the likelihood-based model is weighted by observed commuting flows and therefore gives greater influence to high-volume distance bins. The purpose of the likelihood-based validation is therefore not to reproduce the exact OLS slope value, but to test whether the same negative distance-decay structure remains visible under a different estimation logic. The results support this interpretation.

The model-comparison results further clarify the functional form of the relationship. For all 12 centers, the truncated power-law model is selected by AIC. For 11 of the 12 centers, it is also selected by BIC, with C3 being the only case where the pure power-law specification is preferred by BIC. This pattern suggests that the relevant empirical regularity is better interpreted as bounded opportunity-conditioned distance decay rather than as an exact unbounded power law. This distinction is important because employment-centered commuting fields are spatially limited by the urban boundary, the distribution of residential opportunities, and overlap with neighboring centers.

The goodness-of-fit measures also support a scale-sensitive interpretation. The correlation between the observed commuting share and the model-implied commuting share is high for most centers, ranging from 0.793 to 0.991. Eleven of the 12 centers have correlations above 0.90, and eight exceed 0.97. Centers such as C3, C6, C10, and C9 show especially strong agreement between observed and model-implied distributions. C4 is the weakest case by this metric, with a correlation of 0.793, which is consistent with its more peripheral position and the greater boundedness of its matching field. These weaker cases should not be treated as failures of the framework. Rather, they indicate that some localized employment fields are more spatially directional, more bounded, or more strongly affected by overlap with other centers.

Appendix Table A6 further reports metropolitan-scale likelihood-based validation results for the New York and Chicago MSAs. Both cases show strong negative OLS slopes, high correlations between observed and model-implied commuting shares, and the same AIC-selected truncated power-law specification. These results provide supplementary evidence that bounded opportunity-conditioned distance decay is also visible in large U.S. metropolitan areas at the metropolitan scale.

Taken together, the likelihood-based validation supports the main conclusion of the London decomposition. The opportunity-normalized distance-decay relationship is recurrent across employment-centered subsystems, but it is not identical across them. Its visibility and functional form depend on the spatial scale and internal structure of the metropolitan system. The London case therefore strengthens the paper's central argument: in complex polycentric cities, the relevant empirical unit may be the localized matching field rather than the aggregate citywide boundary.

**Table 1. Likelihood-based validation of bounded distance-decay patterns in London employment-centered subsystems**

| Center | Flow | OLS slope | OLS $R^2$ | MLE-implied slope | KS | Corr. f and q | Best AIC model | Best BIC model |
|---|---|---|---|---|---|---|---|---|
| C1 | 31,272 | −0.868 | 0.803 | −0.728 | 0.022 | 0.979 | Truncated power-law | Truncated power-law |
| C2 | 31,731 | −0.955 | 0.903 | −0.973 | 0.049 | 0.980 | Truncated power-law | Truncated power-law |
| C3 | 22,900 | −0.700 | 0.812 | −0.779 | 0.012 | 0.991 | Truncated power-law | Power-law |
| C4 | 16,587 | −2.126 | 0.895 | −1.833 | 0.156 | 0.793 | Truncated power-law | Truncated power-law |
| C5 | 23,847 | −0.837 | 0.938 | −0.790 | 0.031 | 0.973 | Truncated power-law | Truncated power-law |
| C6 | 21,658 | −0.895 | 0.920 | −0.782 | 0.021 | 0.987 | Truncated power-law | Truncated power-law |
| C7 | 14,965 | −0.854 | 0.934 | −0.815 | 0.040 | 0.969 | Truncated power-law | Truncated power-law |
| C8 | 14,289 | −1.003 | 0.825 | −0.919 | 0.027 | 0.974 | Truncated power-law | Truncated power-law |
| C9 | 12,842 | −0.915 | 0.953 | −0.895 | 0.026 | 0.980 | Truncated power-law | Truncated power-law |
| C10 | 12,940 | −0.893 | 0.926 | −0.892 | 0.042 | 0.980 | Truncated power-law | Truncated power-law |
| C11 | 11,284 | −1.585 | 0.858 | −1.571 | 0.064 | 0.956 | Truncated power-law | Truncated power-law |
| C12 | 9,560 | −1.666 | 0.923 | −1.322 | 0.036 | 0.964 | Truncated power-law | Truncated power-law |

Notes: Flow denotes the observed commuting flow associated with each employment-centered subsystem. The MLE-implied slope, KS statistic, and Corr. f and q are reported for the pure opportunity-conditioned power-law specification. The best model is selected by AIC and BIC from the opportunity-only, pure power-law, exponential, and truncated power-law specifications. KS denotes the Kolmogorov–Smirnov statistic between the observed commuting share and the model-implied commuting share. Corr. f and q denotes the correlation between $f(r)$ and $q(r)$. OLS values are reported using the same log–log diagnostic specification as in Appendix Figure A6.

# 6. Discussion

## 6.1 What opportunity-normalized matching reveals about urban structure

This paper set out to examine whether residence–workplace matching contains an empirical structure that becomes visible only after the city's own opportunity composition is taken into account. The results show that it does. Across the British cities examined here, observed commuting is systematically more concentrated at shorter distances than the opportunity-based pair distribution implied by the joint spatial distribution of residents and jobs. This means that commuting structure cannot be read directly from urban opportunity geometry alone. Cities do not merely contain potential residence–workplace pairings. They also realize some distance-specific pairings more strongly than others.

This distinction clarifies what commuting can reveal about urban spatial structure. A city is not organized only through the location of homes and jobs. It is also organized through the recurrent realization of links between them. Established indicators such as jobs–housing balance, accessibility, average commuting distance, and excess commuting each capture important aspects of urban organization, but they do not directly isolate how the distance structure of potential residence–workplace pairings differs from that of actually realized pairings. The opportunity-normalized framework developed here makes this relation explicit.

The main contribution of this distinction is not simply that short-distance commuting is common. That fact is already well known. The more important finding is that short-distance matching remains disproportionately strong even after the underlying opportunity field is accounted for. In other words, realized residence–workplace matching contains a distance-sensitive structure that is not mechanically implied by the spatial distribution of homes and jobs alone. This structure provides a useful way to describe how cities translate potential urban opportunity into actual home–work connections.

### 6.2 From a commuting indicator to a parameter of realized matching

The estimated distance-decay coefficient provides a compact summary of this realized matching structure. It describes how sharply opportunity-normalized residence–workplace matching attenuates with distance within a given analytical unit. Its value lies in linking two layers of urban structure: the opportunity field generated by the distribution of residences and employment, and the realized matching field observed through commuting flows.

This interpretation distinguishes the coefficient from more conventional commuting indicators. Average commuting distance summarizes realized outcomes, but it does not show whether those outcomes are disproportionate relative to the city's own opportunity structure. Jobs–housing balance describes the spatial arrangement of residents and jobs, but it does not identify which distance-specific pairings are actually realized. Excess-commuting measures compare observed commuting with optimized or counterfactual allocations, but they are primarily concerned with benchmarked efficiency. By contrast, the opportunity-normalized coefficient summarizes the distance-sensitive intensity with which potential residence–workplace pairings are transformed into realized connections.

For this reason, the coefficient is best understood as a descriptive structural summary of realized urban matching rather than as a fully identified mechanism. It links the opportunity field generated by the distribution of residences and employment with the realized matching field observed through commuting flows. Its interpretation should therefore remain connected to the broader spatial, market, transport, and institutional conditions under which home–work connections are formed. This makes it useful not as a final behavioral explanation, but as a parsimonious representation of how urban systems organize realized home–work connections relative to their own opportunity composition.

### 6.3 Scale-sensitive regularity and the meaning of the London case

The London results show that the relevant scale of this parameter cannot be assumed in advance. In relatively compact cities, the citywide unit may approximate a dominant residence–workplace matching field. In more internally differentiated metropolitan systems, however, the citywide curve may combine several partially distinct matching fields into one aggregate profile. In such cases, a citywide coefficient may remain descriptively useful, but it may not correspond to a single coherent matching structure.

This is the central significance of the London decomposition. At the citywide scale, London displays a declining but flatter and more compositionally complex opportunity-normalized pattern than Birmingham and Manchester. After decomposition into 12 employment-centered subsystems, negative distance-decay relationships are recovered across all centers, although their slopes and goodness of fit differ substantially. This suggests that London's aggregate pattern is not simply an exception to the broader regularity. Rather, it reflects the superposition of multiple localized residence–workplace systems.

The heterogeneous center-level results in London should therefore not be interpreted as evidence that the proposed framework fails in large metropolitan systems. Rather, they clarify the conditions under which the coefficient is most interpretable. A localized employment center does not necessarily define an independent and self-contained residence–workplace matching field. In a strongly hierarchical metropolis, secondary centers may remain embedded within the wider influence of dominant core areas, transport corridors, uneven residential opportunity distributions, and broader amenity structures. Their observed matching patterns may therefore reflect the overlap of several spatial fields rather than distance from the local center alone. From this perspective, weaker or more curved center-level fits are not necessarily evidence of absent distance sensitivity. They may instead indicate that the chosen subsystem only partially approximates a coherent matching field.

The implication is that heterogeneity in opportunity-normalized matching has two sources. One is genuine cross-city difference in how urban systems organize home–work connections. The other is scale mismatch between the analytical boundary and the internally coherent matching field. This distinction matters because it shifts the interpretation of the coefficient from a simple city-level statistic toward a scale-contingent measure of realized residence–workplace organization. In complex metropolitan areas, the most meaningful analytical unit may be an employment-centered matching field rather than the city as a whole. The London case therefore does not weaken the meaning of the parameter. Instead, it shows that the parameter's interpretability depends on whether the analytical unit approximates a coherent residence–workplace matching field.

### 6.4 Implications for urban structure modeling

The findings have implications for how urban structure may be modeled. Many urban models describe the spatial distribution of residents, jobs, centers, density, accessibility, or land use. These dimensions are essential, but they mainly describe spatial supply and potential opportunity. The present results suggest that urban structure also includes a realized matching layer: the way homes and workplaces are actually connected within the opportunity field of the city.

A parameter summarizing opportunity-normalized matching could therefore help connect urban morphology with everyday urban functioning. It provides a way to move from the location of residences and jobs to the realized relations between them. In future models, this kind of parameter could be used to represent how strongly a given urban system organizes home–work matching toward shorter distances after accounting for its own opportunity structure. This would be especially relevant for models concerned with the relation between employment centers, residential distributions, and daily mobility.

The London results further suggest that future urban structure models should not treat the citywide scale as the only possible level of parameterization. In simpler urban systems, a city-level matching parameter may be a reasonable approximation. In larger and more polycentric systems, however, center-based or subsystem-based parameters may better represent the internal organization of realized matching. This scale-sensitive perspective may help future models describe not only the distribution of urban opportunities, but also the spatial level at which those opportunities are translated into realized residence–workplace connections.

### 6.5 Planning relevance: from spatial supply to realized spatial needs

The framework also has planning relevance. Urban planning often focuses on the spatial supply side of cities: where housing is located, where employment is concentrated, where transport infrastructure is provided, and how accessible opportunities appear to be. These are necessary dimensions of urban evaluation, but they do not fully reveal how residents actually connect living and working locations within that supplied structure.

Opportunity-normalized matching provides an additional perspective. It asks whether realized home–work connections are organized differently from what the city's opportunity geometry alone would imply. This is important because a city may contain many jobs and housing units, but still connect them in ways that are spatially burdensome, uneven, or poorly aligned with residents' actual matching patterns. The coefficient identified in this paper can therefore be read as a summary of how the urban system translates spatial supply into realized residence–workplace integration.

From a planning perspective, this shifts attention from the presence of opportunities to the realized relation among opportunities. It encourages evaluation of whether the spatial arrangement of jobs, housing, and transport supports workable connections between where people live and where they work. It also provides a possible empirical reference for comparing urban systems, identifying cases where realized matching is more spatially constrained, and examining whether complex metropolitan structures require planning at the level of subcenters or localized matching fields rather than at the citywide scale alone.

This interpretation should not be taken to mean that shorter matching is always normatively better. A steep distance-decay relationship may reflect compact integration, but it may also reflect constraints that limit access to more distant jobs. A flatter relationship may reflect broader labor-market integration, but it may also indicate long commuting burdens. The value of the framework is therefore diagnostic rather than prescriptive: it reveals how residence–workplace matching is organized relative to opportunity, while normative interpretation should be assessed alongside housing affordability, transport quality, labor-market access, and distributional equity.

### 6.6 Limitations and future research

Several limitations should be kept in view. First, the opportunity baseline is deliberately parsimonious. It is constructed from the joint spatial distribution of residents and jobs, and it does not incorporate occupational compatibility, income, housing tenure, transport mode, or individualized travel costs. This makes the framework transparent and comparable across cities, but it also means that the normalized pattern reflects a broad combination of spatial, social, market, and transport-related forces.

Second, the analysis uses straight-line centroid distance rather than network distance or travel time. The results therefore describe attenuation over geometric separation rather than generalized commuting cost. This distinction may matter especially in large and complex metropolitan systems, where transport networks, physical barriers, and mode availability can create substantial differences between Euclidean distance and experienced commuting cost.

Third, the main empirical evidence is based on 2021 Output Area-level data for British cities. The supplementary U.S. metropolitan results provide external validation, but the paper should still be understood as a British city-based analysis with additional cross-national evidence rather than as a comprehensive cross-national comparison. Because the British data come from 2021, the estimated matching patterns may also reflect commuting conditions during a period affected by pandemic-related changes in work arrangements. The results should therefore be interpreted as evidence on the observed residence–workplace matching structure in that data context.

Fourth, the London decomposition identifies employment-centered subsystems using high-employment Output Areas, peak separation, and assignment radius rules. Although all 12 employment-centered subsystems have sufficient observed commuting flow and fitted distance-bin support for the present analysis, the resulting center boundaries should not be interpreted as definitive functional labor-market boundaries. Some localized matching fields may still overlap with nearby centers or be shaped by directional transport corridors and uneven residential opportunity distributions. The decomposition should therefore be interpreted as an empirical approximation of localized matching fields rather than a complete partition of London's commuting system. This reinforces the paper's scale-sensitive argument, but it also shows that coherent matching fields must be empirically supported rather than assumed from center locations alone.

Fifth, the present analysis does not identify the mechanisms through which specific employment centers generate particular residential matching fields. Residence–workplace matching differs from many forms of same-type mobility because it connects heterogeneous residential and employment locations. A single employment center may contain workers with different incomes, occupations, skills, housing constraints, and commuting tolerances. These groups may face different feasible residential fields and may weigh workplace distance, housing conditions, local amenities, school access, transport services, and proximity to larger urban cores in different ways. The center-level curves estimated here therefore represent aggregate matching fields, not group-specific mechanisms of residential choice.

These limitations point to several directions for future research. One priority is to test whether the same opportunity-normalized regularity remains visible when distance is measured by network distance or travel time. A second is to construct more differentiated opportunity baselines by occupation, skill group, income, housing tenure, or transport mode. Future research could estimate group-specific opportunity baselines and group-specific normalized matching intensities to examine whether apparently unstable aggregate center-level curves reflect the mixture of multiple socioeconomic matching fields. A third direction is to examine how the coefficient varies with metropolitan size, employment concentration, polycentricity, transport infrastructure, housing-market conditions, and amenity distributions. A fourth is to extend the framework more systematically to additional countries and data environments to assess whether the distinction between opportunity field and realized matching remains useful under different institutional and spatial contexts.

Overall, the paper suggests that residence–workplace matching should be understood not only as a commuting outcome, but also as a structured relation between what an urban system makes possible and what it realizes. Once the opportunity field is taken into account, realized home–work matching exhibits a recurrent but heterogeneous distance-decay structure. Interpreted carefully, this structure provides a step toward a more explicit parameterization of realized residence–workplace organization within urban spatial analysis.

# 7. Conclusion

This paper examined whether realized residence–workplace matching contains a distance-sensitive structure that is distinct from the opportunity geometry of the city itself. Using Output Area-level data for British cities, it compared observed within-city commuting distributions with opportunity-based pair distributions constructed from all residence–workplace combinations weighted by residential and employment mass. The ratio between these two distributions was interpreted as an opportunity-normalized realized matching intensity.

The analysis produced three main findings. First, observed commuting is systematically more concentrated at shorter distances than the opportunity-based pair distribution. This shows that realized

residence–workplace matching cannot be understood as a direct reflection of the spatial distribution of homes and jobs alone. Cities selectively realize some distance-specific residence–workplace pairings more strongly than others.

Second, after opportunity normalization, realized matching often declines with distance in a clear and comparable way. In many cases, this relationship is approximately linear in log–log space and can be summarized by an empirical distance-decay coefficient. Its value lies in providing a compact description of how an urban system transforms potential residence–workplace pairings into realized home–work connections, while leaving room for variation across cities, scales, and local matching fields.

Third, the London case shows that the visibility of this regularity is scale-sensitive. At the citywide scale, London appears flatter and more compositionally complex than the more compact comparison cities. After decomposition into 12 employment-centered subsystems, negative localized matching relationships are consistently recovered, although their steepness and fit vary across centers. This indicates that aggregate deviation may reflect the superposition of multiple matching fields rather than the absence of an underlying structure. Although these U.S. cases are used only as supplementary validation rather than as a full cross-national comparison, they show that the same bounded opportunity-normalized attenuation pattern can also be observed in large metropolitan areas outside Britain.

These findings contribute to the study of commuting and urban spatial structure by identifying opportunity-normalized residence–workplace matching as a distinct analytical object. The paper extends the broader logic of separating opportunity structure from realized movement into the domain of home–work matching. It also shows that the meaningful unit for interpreting this structure may not always be the whole city. In complex metropolitan systems, the coherent matching field may provide a more appropriate scale for understanding how urban space connects residences and workplaces.

More broadly, the paper argues that urban spatial structure should be understood not only as a configuration of homes, jobs, and infrastructure, but also as a pattern of selective realization linking them. The contribution of the paper lies in making this realized matching layer visible and measurable. Future research can build on this framework by incorporating travel time, occupational matching, transport mode, housing-market segmentation, and more systematic cross-national comparisons. Such extensions would help clarify how opportunity-normalized matching varies across different urban systems and how it can inform more realistic models of urban structure and planning.

# Appendix

**Appendix Table A1. Descriptive statistics for the 12 London employment-centered subsystems**

| Center | Cluster total jobs | Peak jobs | High-employment OAs in cluster | Observed commuting flow | Unique residential origins | Positive fitted distance bins | Distance range (km) | Minimum bin flow | Maximum member distance from peak (m) |
|---|---|---|---|---|---|---|---|---|---|
| C1 | 42,959 | 33,025 | 4 | 31,272 | 15,075 | 59 | 0.75–30.75 | 1 | 760.8 |
| C2 | 36,878 | 5,734 | 10 | 31,731 | 14,903 | 60 | 0.75–30.25 | 1 | 814.3 |
| C3 | 31,800 | 12,590 | 5 | 22,900 | 12,365 | 59 | 0.75–29.75 | 1 | 880.9 |
| C4 | 31,378 | 18,894 | 3 | 16,587 | 4,785 | 102 | 0.75–51.75 | 1 | 683.4 |
| C5 | 30,470 | 10,527 | 6 | 23,847 | 12,983 | 55 | 0.75–27.75 | 2 | 848.3 |
| C6 | 27,761 | 9,137 | 5 | 21,658 | 11,903 | 57 | 0.75–29.25 | 2 | 891.4 |
| C7 | 18,465 | 4,328 | 7 | 14,965 | 9,502 | 56 | 0.75–28.25 | 1 | 835.5 |
| C8 | 18,333 | 9,980 | 3 | 14,289 | 8,354 | 66 | 0.75–34.75 | 1 | 495.7 |
| C9 | 15,773 | 6,656 | 5 | 12,842 | 8,399 | 57 | 0.75–28.75 | 2 | 589.3 |
| C10 | 14,980 | 5,328 | 4 | 12,940 | 8,424 | 60 | 0.75–30.25 | 1 | 356.5 |
| C11 | 13,063 | 7,124 | 3 | 11,284 | 4,872 | 73 | 0.75–38.75 | 1 | 570.7 |
| C12 | 10,869 | 8,723 | 2 | 9,560 | 5,118 | 56 | 0.75–28.25 | 2 | 508.6 |

Notes: Cluster total jobs reports the total number of jobs within each employment-center cluster. Peak jobs denote the number of jobs in the peak Output Area of each center. Observed commuting flow refers to the total observed commuting volume associated with each employment-centered subsystem. Unique residential origins denote the number of residential Output Areas contributing observed commuting flows. Positive fitted distance bins refer to the distance bins included in the log–log and likelihood-based fitting support. Distance range is reported using bin-center distances.

**Appendix Table A2. Identification settings for London employment centers**

| Item | Value | Purpose |
|---|---|---|
| High-employment OA definition | Top 0.5% jobs OAs | Identify the upper tail of employment concentration |
| Peak separation distance | 1,200 m | Separate nearby employment peaks into distinct employment centers |
| Assignment radius | 1,600 m | Assign nearby high-employment OAs to employment centers |
| Number of employment centers | 12 | Define employment-centered subsystems for London decomposition |
| Distance bin width | 500 m | Construct binned observed and opportunity-based distance distributions |
| Minimum starting distance | 0.6 km | Exclude the immediate zero-distance and same-location range |

Notes: This table summarizes the main parameter settings used to identify employment-centered subsystems in London. High-employment Output Areas are defined from the upper tail of the employment distribution. Nearby peaks are separated using the peak separation distance, and surrounding high-employment OAs are assigned to each employment center within the assignment radius.

**Appendix Table A3. Robustness of center-level estimates under minimum bin-flow thresholds**

| Min. bin flow | Centers | Retained centers | Mean OLS slope | Median OLS slope | Mean OLS R² | Minimum OLS R² | Mean MLE slope | Mean KS | Mean Corr. f and q | AIC TPL | BIC TPL |
|---|---|---|---|---|---|---|---|---|---|---|---|
| 5 | 12 | C1-C12 | −1.106 | −0.904 | 0.931 | 0.900 | −1.023 | 0.044 | 0.958 | 12 | 11 |
| 10 | 12 | C1-C12 | −1.098 | −0.916 | 0.934 | 0.890 | −1.020 | 0.044 | 0.956 | 12 | 11 |
| 20 | 12 | C1-C12 | −1.080 | −0.927 | 0.943 | 0.875 | −1.015 | 0.044 | 0.954 | 11 | 10 |

Notes: This robustness check restricts the fitted support to consecutive distance bins satisfying alternative minimum observed-flow thresholds. The full set of 12 employment-centered subsystems is retained under all three thresholds. OLS slopes and OLS $R^2$ are calculated from the log–log relationship between opportunity-normalized matching intensity and distance. The MLE-implied slope is obtained from the opportunity-conditioned power-law specification. KS denotes the Kolmogorov–Smirnov statistic between the observed commuting share and the model-implied commuting share. Corr. f and q denotes the correlation between $f(r)$ and $q(r)$. The final two columns report the number of centers for which the truncated power-law model is selected by AIC or BIC. TPL denotes truncated power-law.

**Appendix Table A4. Summary of model-selection results for London employment-centered subsystems**

| Model | Number selected by AIC | Number selected by BIC |
|---|---|---|
| Opportunity only | 0 | 0 |
| Power-law | 0 | 1 |
| Exponential | 0 | 0 |
| Truncated power-law | 12 | 11 |
| Total | 12 | 12 |

Notes: Model selection is based on four candidate specifications: opportunity-only, pure power-law, exponential decay, and truncated power-law. The opportunity-only model is defined as $q(r) = p(r)$. The pure power-law model is defined as $q(r; s) \propto p(r)r^{-s}$. The exponential model is defined as $q(r; \lambda) \propto p(r)e^{-\lambda r}$. The truncated power-law model is defined as $q(r; s, \lambda) \propto p(r)r^{-s}e^{-\lambda r}$. AIC and BIC denote Akaike and Bayesian information criteria, respectively.

**Appendix Table A5. Empirical support of binned power-law estimation in the 12 London employment-centered subsystems**

| Center | $n_{bins}$ | $x_{min}(m)$ | $x_{max}(m)$ | Observed commuting flow | Exponent | MLE-implied slope | KS | Corr. f and q |
|---|---|---|---|---|---|---|---|---|
| C1 | 59 | 750 | 30,750 | 31,272 | 0.728 | −0.728 | 0.022 | 0.979 |
| C2 | 60 | 750 | 30,250 | 31,731 | 0.973 | −0.973 | 0.049 | 0.980 |
| C3 | 59 | 750 | 29,750 | 22,900 | 0.779 | −0.779 | 0.012 | 0.991 |
| C4 | 102 | 750 | 51,750 | 16,587 | 1.833 | −1.833 | 0.156 | 0.793 |
| C5 | 55 | 750 | 27,750 | 23,847 | 0.790 | −0.790 | 0.031 | 0.973 |
| C6 | 57 | 750 | 29,250 | 21,658 | 0.782 | −0.782 | 0.021 | 0.987 |
| C7 | 56 | 750 | 28,250 | 14,965 | 0.815 | −0.815 | 0.040 | 0.969 |
| C8 | 66 | 750 | 34,750 | 14,289 | 0.919 | −0.919 | 0.027 | 0.974 |
| C9 | 57 | 750 | 28,750 | 12,842 | 0.895 | −0.895 | 0.026 | 0.980 |
| C10 | 60 | 750 | 30,250 | 12,940 | 0.892 | −0.892 | 0.042 | 0.980 |
| C11 | 73 | 750 | 38,750 | 11,284 | 1.571 | −1.571 | 0.064 | 0.956 |
| C12 | 56 | 750 | 28,250 | 9,560 | 1.322 | −1.322 | 0.036 | 0.964 |

*Notes: $x_{min}$ and $x_{max}$ denote the minimum and maximum bin-center distances used in the likelihood-based power-law estimation and are reported in meters. Because the estimation is based on binned distance distributions rather than individual OD-level distances, $n_{bins}$ denotes the number of positive distance bins included in the fitted support. Observed commuting flow reports the total observed commuting volume covered by the same fitted support. The opportunity-conditioned power-law model is defined as $q(r;s) \propto p(r)r^{-s}$, where $p(r)$ is the opportunity-based pair distribution. The exponent corresponds to $s$, and the MLE-implied slope is reported as $-s$. KS denotes the Kolmogorov–Smirnov statistic between the observed commuting share $f(r)$ and the model-implied commuting share $q(r)$. Corr. f and q denotes the correlation between $f(r)$ and $q(r)$.*

**Appendix Table A6. Metropolitan-scale likelihood validation in U.S. metropolitan areas**

| Metropolitan area | n bins | Distance range (km) | Observed flow | OLS slope | OLS $R^2$ | MLE-implied slope | Corr. f and q | Best AIC model |
|---|---|---|---|---|---|---|---|---|
| New York MSA | 199 | 1.00–100.00 | 6,788,645 | −1.177 | 0.980 | −0.985 | 0.970 | TPL |
| Chicago MSA | 199 | 1.00–100.00 | 3,638,200 | −1.188 | 0.963 | −0.959 | 0.950 | TPL |

*Notes: U.S. results use 2022 block-level LODES data, Census-defined metropolitan boundaries, 500-meter distance bins. Observed flow refers to commuting volume included in the fitted support. Corr. f and q denotes the correlation between observed and model-implied commuting shares. AIC denotes Akaike information criterion.*

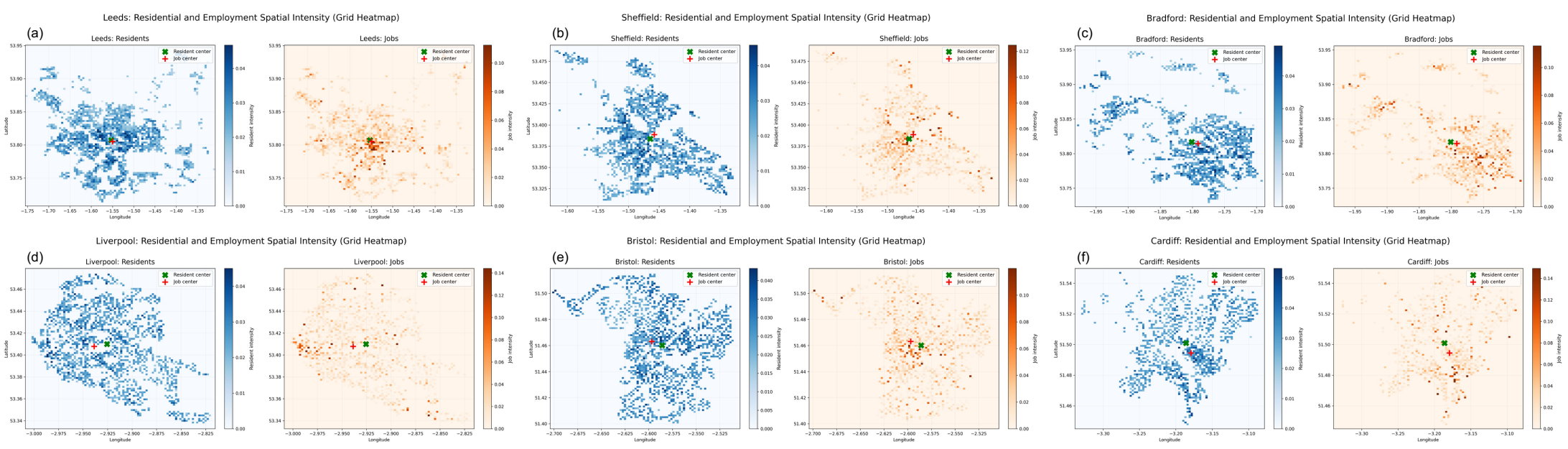

**Appendix Figure A1. Residential and employment spatial intensity in supplementary British cities**

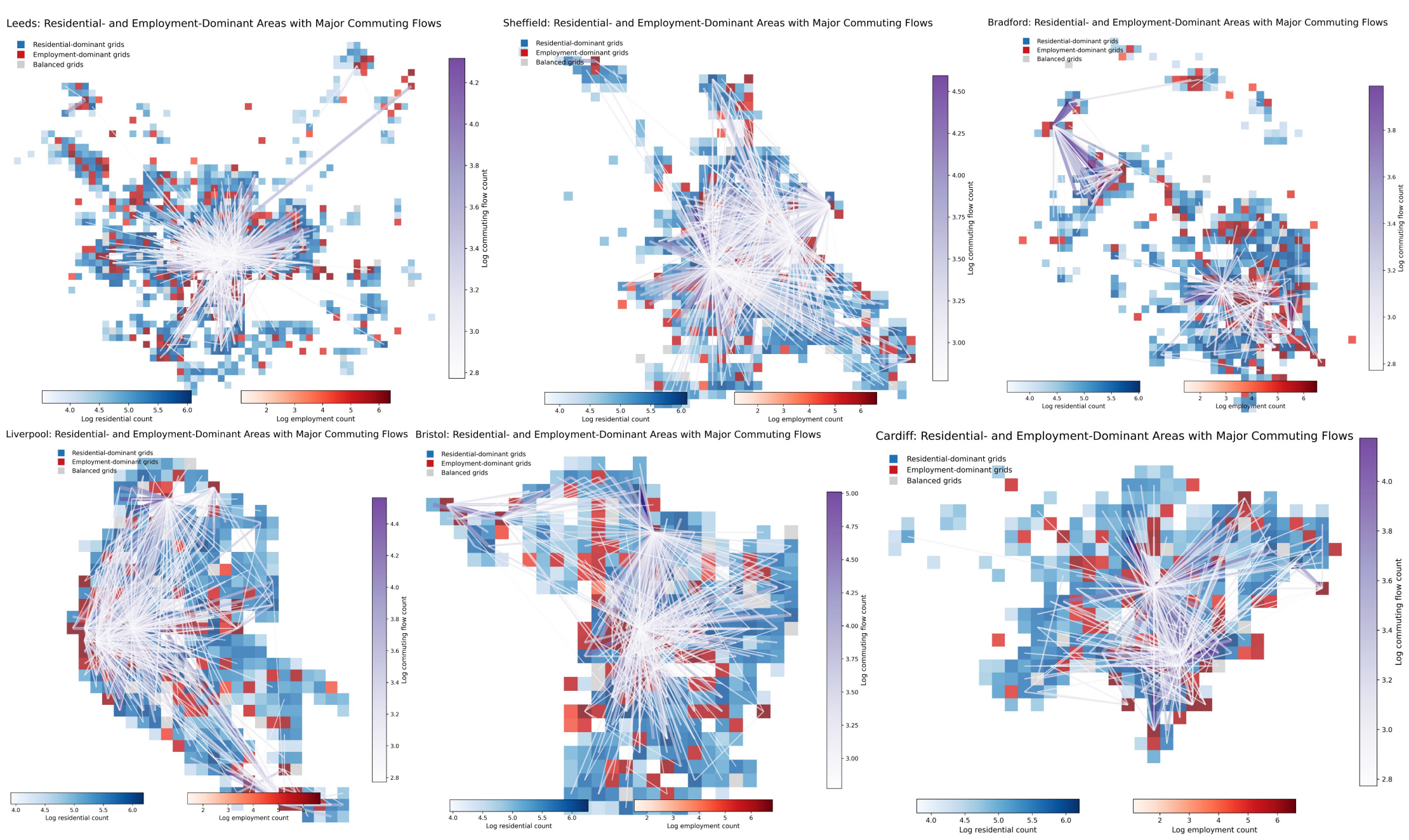

**Appendix Figure A2. Residential- and employment-dominant areas with major commuting flows in supplementary British cities**

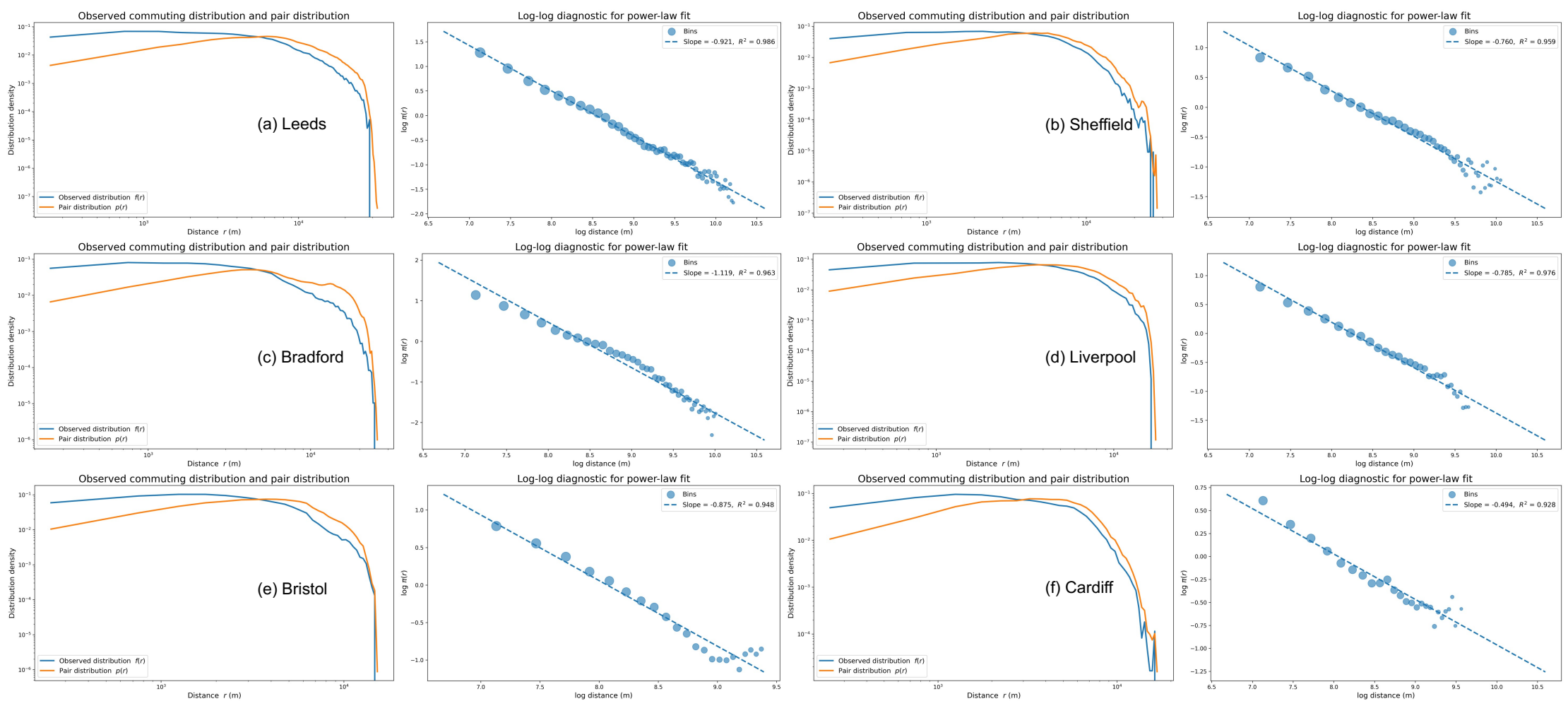

**Appendix Figure A3. Observed commuting distributions, opportunity-based pair distributions, and opportunity-normalized realized matching intensity in supplementary British cities**

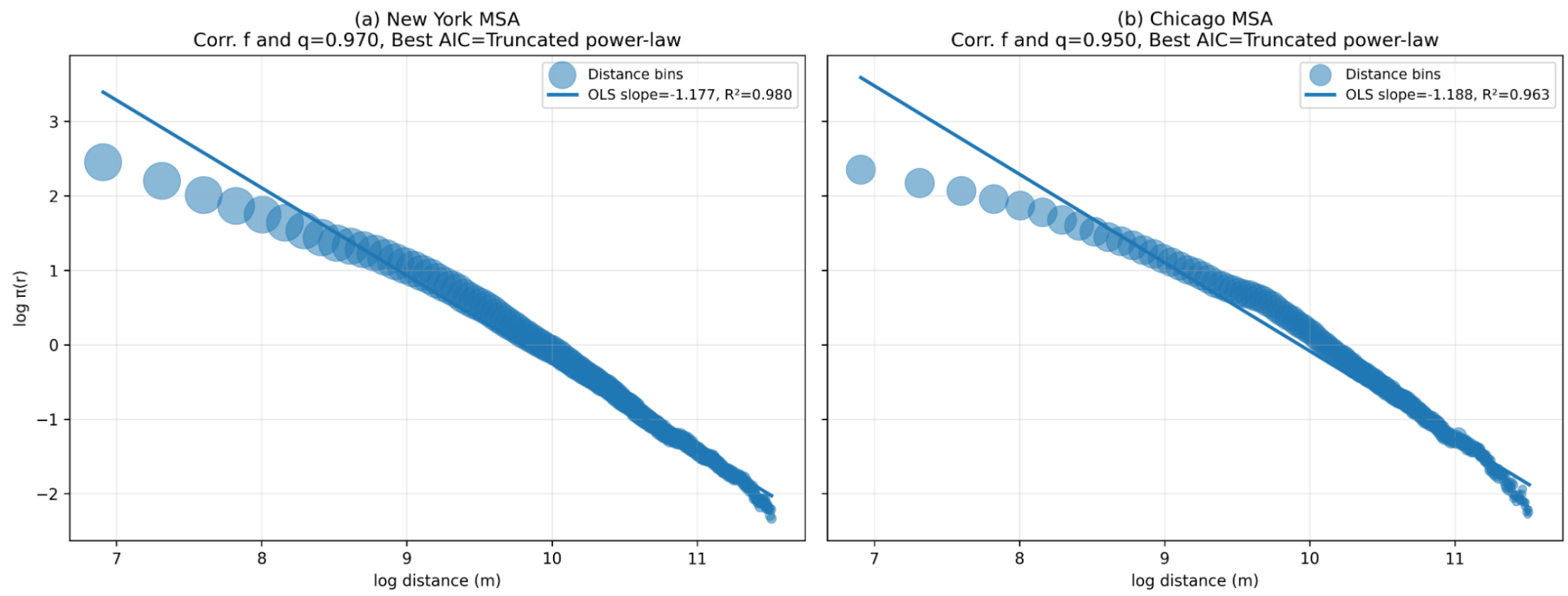

**Appendix Figure A4. Opportunity-normalized matching intensity in U.S. metropolitan areas**

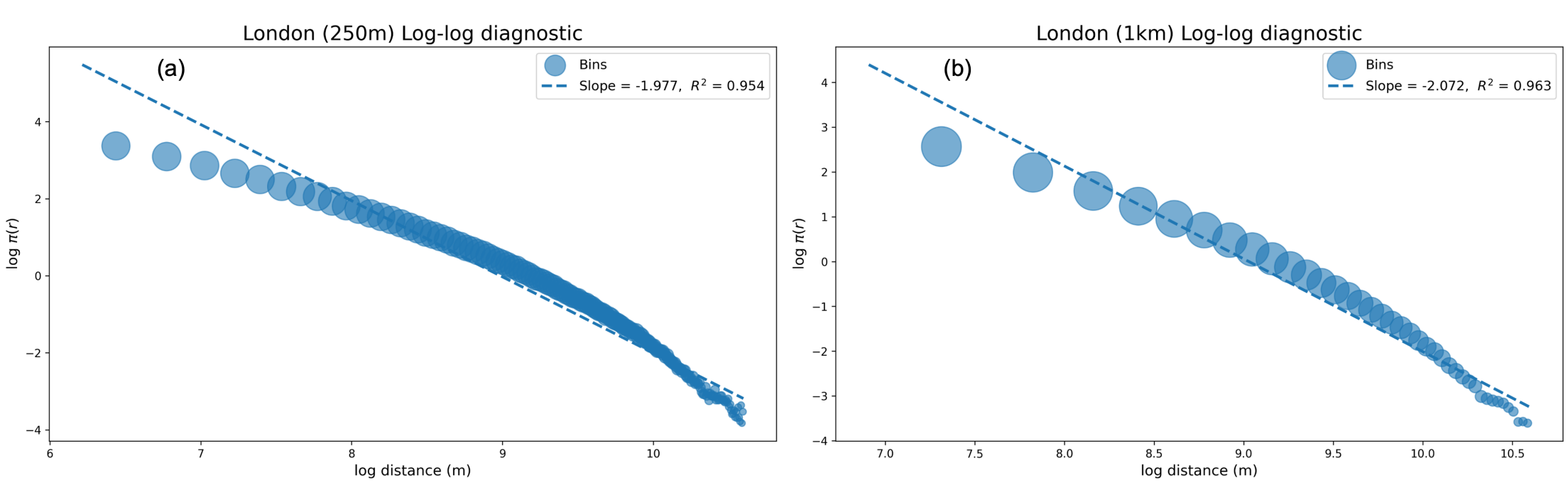

**Appendix Figure A5. Log–Log diagnostics for opportunity-normalized realized matching in London under alternative distance bin widths**

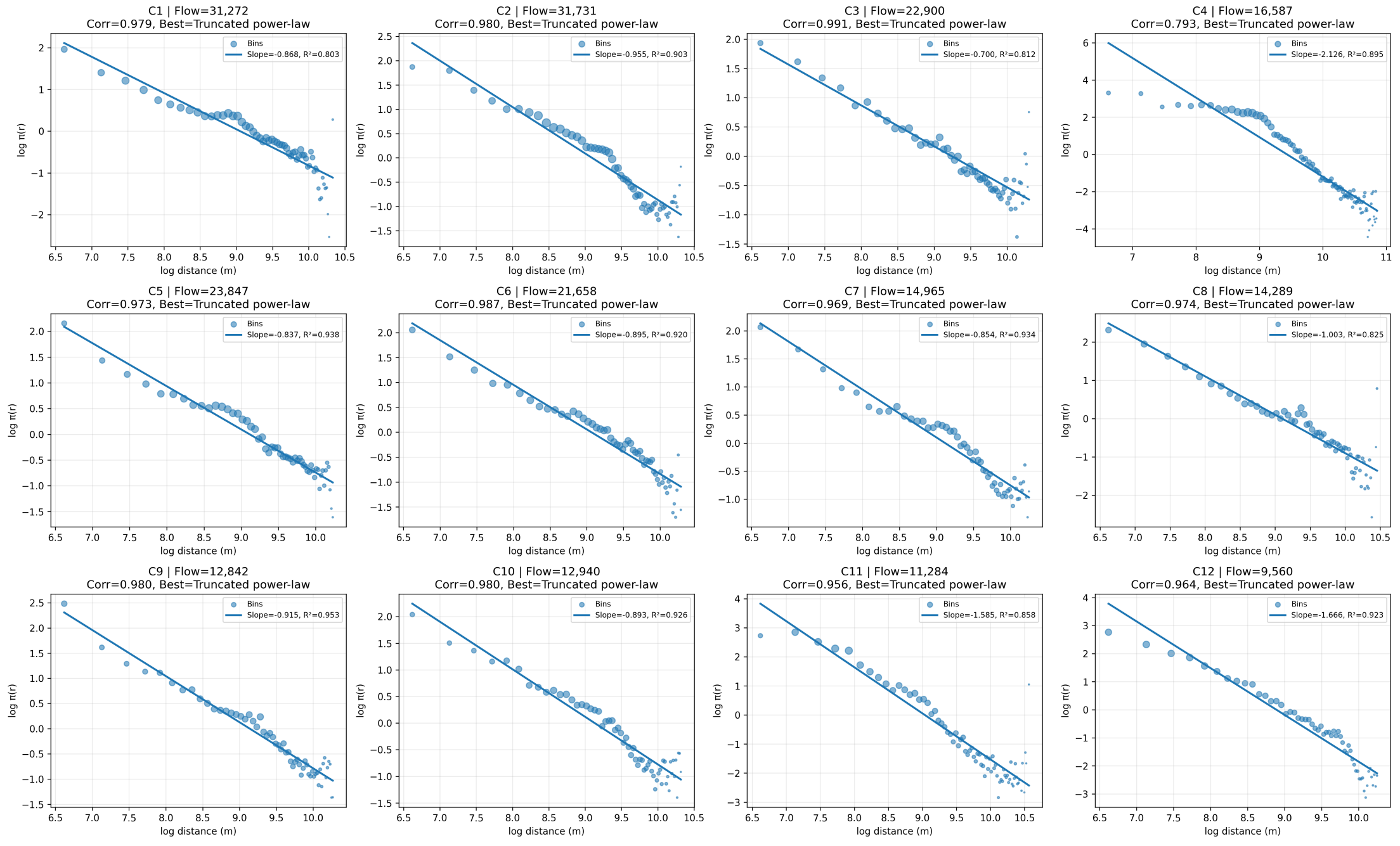

**Appendix Figure A6. Center-specific log–log diagnostics for London employment-centered subsystems**

## Data availability

The UK origin–destination commuting data used in this study were obtained from the 2021 Census origin–destination data provided by Nomis: https://www.nomisweb.co.uk/sources/census_2021_od. The U.S. origin–destination employment statistics were obtained from the Longitudinal Employer-Household Dynamics (LEHD) data program of the U.S. Census Bureau: https://lehd.ces.census.gov/data/. The processed data generated for this study are available from the corresponding author upon reasonable request.